\documentclass[aps,pre,10pt,superscriptaddress,nofootinbib,onecolumn,amsmath,showpacs,amssymb,final]{revtex4-1}
\usepackage{graphicx}
\usepackage{bm}
\usepackage[utf8]{inputenc}
\usepackage[colorlinks=true,linkcolor=blue,urlcolor=blue,citecolor=blue,anchorcolor=blue]{hyperref}
\newcommand{\T}{\intercal}
\usepackage[caption=false]{subfig}

\begin{document}

\title{Universality of the stochastic block model}

\author{Jean-Gabriel Young}
\email{jean-gabriel.young.1@ulaval.ca}
\affiliation{D\'epartement de physique, de g\'enie physique et d'optique, Universit\'e Laval, Qu\'ebec (Qu{\'e}bec), Canada G1V 0A6}
\affiliation{Centre interdisciplinaire de mod\'elisation math\'ematique de l'Universit\'e Laval, Qu\'ebec (QC), G1V 0A6, Canada}

\author{Guillaume St-Onge}
\affiliation{D\'epartement de physique, de g\'enie physique et d'optique, Universit\'e Laval, Qu\'ebec (Qu{\'e}bec), Canada G1V 0A6}
\affiliation{Centre interdisciplinaire de mod\'elisation math\'ematique de l'Universit\'e Laval, Qu\'ebec (QC), G1V 0A6, Canada}

\author{Patrick Desrosiers}
\affiliation{D\'epartement de physique, de g\'enie physique et d'optique, Universit\'e Laval, Qu\'ebec (Qu{\'e}bec), Canada G1V 0A6}
\affiliation{Centre interdisciplinaire de mod\'elisation math\'ematique de l'Universit\'e Laval, Qu\'ebec (QC), G1V 0A6, Canada}
\affiliation{Centre de recherche CERVO, Qu\'ebec (QC), G1J 2G3, Canada}

\author{Louis J. Dub\'e}
\email{ljd@phy.ulaval.ca}
\affiliation{D\'epartement de physique, de g\'enie physique et d'optique, Universit\'e Laval, Qu\'ebec (Qu{\'e}bec), Canada G1V 0A6}
\affiliation{Centre interdisciplinaire de mod\'elisation math\'ematique de l'Universit\'e Laval, Qu\'ebec (QC), G1V 0A6, Canada}

\date{\today}

\begin{abstract}
Mesoscopic pattern extraction (MPE) is the problem of finding a partition of the nodes of a complex network that maximizes some objective function.
Many well-known network inference problems fall in this category, including, for instance, community detection, core-periphery identification, and imperfect graph coloring.
In this paper, we show that the most popular algorithms designed to solve MPE problems can in fact be understood as special cases of the maximum likelihood formulation of the stochastic block model (SBM),  or one of its direct generalizations.
These equivalence relations show that the SBM is nearly universal with respect to MPE problems.
\end{abstract}

\maketitle

\section{Introduction}
\label{sec:introduction}
Whether it is called community detection, graphical inference, spectral embedding, unsupervised learning, bisection or graph coloring, the idea of summarizing the structure of a complex system by grouping its elements in blocks is a popular one, discovered time and time again in different areas of science \cite{Moore2017}.
As such, there are now a plethora of algorithms and techniques---developed essentially in parallel---that provide good solutions to this ubiquitous problem \cite{Fortunato2016}.
In the past few years, a great deal of work has been done toward unifying and contrasting these approaches, building bridges across cultural divides \cite{Moore2017,Schaub2017}.
This has been fruitful work thus far, for---sometimes surprising---equivalences between drastically different methods have turned up in the process, e.g., between  modularity and the maximum likelihood formulation of the degree-corrected stochastic block models (SBM) \cite{Zhang2014b,newman2016equivalence,roxana2018relating,veldt2017unifying}, various spectral methods \cite{Newman2013}, normalized-cut \cite{kawamoto2015detectability}, random-walks \cite{masuda2017random}, and non-negative matrix factorization \cite{chang2018approximate}.
These results invite the question: Is there a deeper reason for the correspondences, or are they simply mathematical coincidences?

The purpose of this paper is to show that equivalences arise because most of these \emph{mesoscopic pattern extraction} (MPE) methods are actually the maximum likelihood formulation of the SBM in disguise (and a generalization of its degree--corrected version \cite{Karrer2011,Ball2011}).
By MPE problems, we mean any problem where one is asked to find a partition of the network that maximizes some implicit or explicit score, encoded via an objective function.

Our results rest on the concepts of  \emph{equivalence} and \emph{specialization} of the objective functions: Two objective functions are equivalent when they order any pair of partitions the same way (i.e., they implement the same notion of optimality), and specialization refers to the idea of limiting the expressiveness of an objective function by fixing some of its parameters (see Sec.~\ref{sec:mpe_problem}).
With these two operations, we delineate a  hierarchy that crystallizes the idea of the SBM as a general MPE tool: Through specialization of its likelihood,  it can be tailored to find patterns such as assortative and disassortative communities \cite{newman2016equivalence}, bipartite structures \cite{larremore2014efficiently}, or core-periphery splits \cite{Borgatti2000} (in Sec.~\ref{sec:hierarchy}).
Importantly, we show that these specialized likelihoods are \emph{exactly} equivalent to the objective functions implemented by MPE algorithms such as modularity maximization, balanced cut, core-periphery search, etc.
Our framework therefore offers principled methods to determine any arbitrary parameters that might arise in otherwise \empty{ad hoc} modularities \cite{newman2016equivalence}, but also suggests statistical techniques to carry out principled inference, in the spirit of Refs.~\cite{sulc2010belief,Zhang2014b} (see Sec.~\ref{sec:discuss}).


\section{Mesoscopic structures and optimization}
\label{sec:mpe_problem}

\begin{figure}
    \centering
    \includegraphics[width=0.45\linewidth]{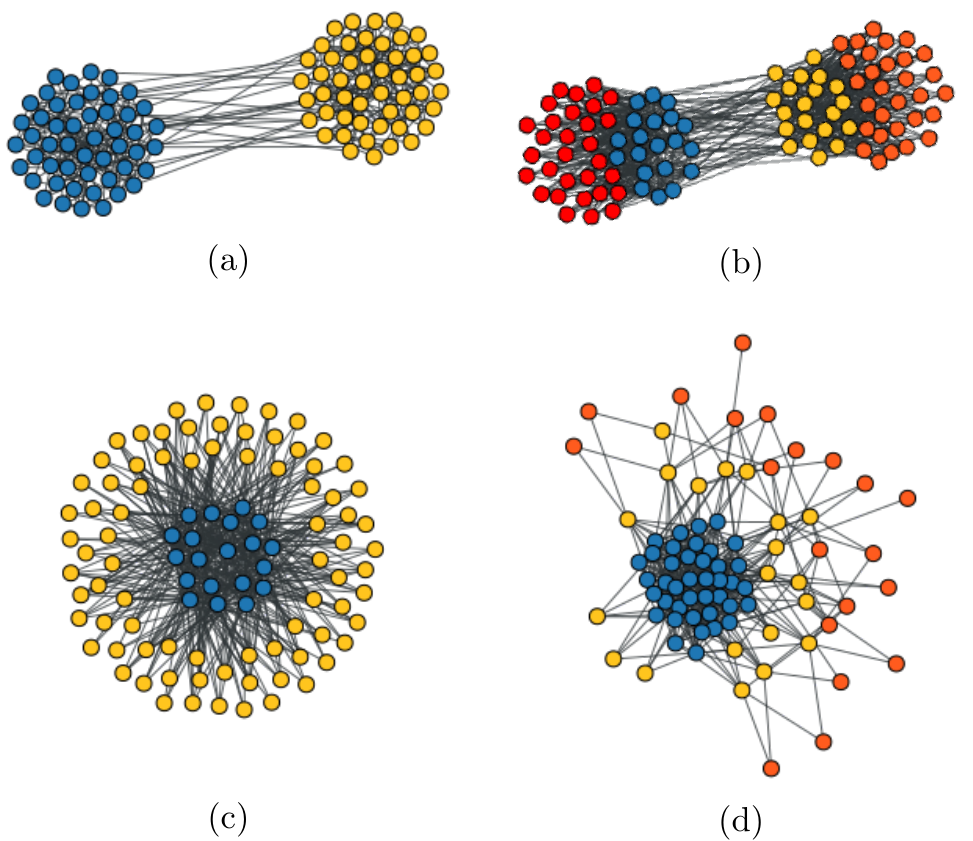}
    \caption{Four examples of mesoscopic pattern extraction problems on artificial networks. (a) Community detection \cite{Fortunato2010,Newman2004}, (b) community detection with further structure, (c) the identification of a simple core-periphery split \cite{Borgatti2000,Rombach2014}, (d) identification of a core with a structured periphery.  
    The targeted patterns are identified with colors.
    }
    \label{fig:mpe}
\end{figure}

The mesoscopic pattern extraction (MPE) problem is usually stated as follows.
We are given an extremely large complex network, generated by some random hidden process.
Its overall organization is impossible to grasp, because its structure is much too detailed.
Our goal with MPE is to reduce this complexity, by subsuming nodes in larger coherent units, using the structure of the network as our only input (and possibly additional metadata \cite{hric2016network,newman2016structure}).
Sometimes, the hope is to reveal functional components and hints about assembly mechanisms, while at other times it is only a matter of making the dataset more manageable, or interpolating from what is known \cite{arenas2007size,clauset2008hierarchical,blondel2008fast,Fortunato2010,Fortunato2016,Porter2009,Schaub2017,shai2017case}.
There is, however, a common theme: MPE algorithms take a complex network as their input, and produce as output a \emph{partition} $\mathcal{B}=\{B_1,...,B_q\}$ of the $n$ nodes in $q$ blocks $B_1,...,B_q$, assigning precisely one block to each node.
Despite these commonalities, the definition of what is a suitable partition will of course depend on the MPE problem at hand; there is thus a wealth of MPE \emph{algorithms}, reflecting the wealth of MPE \emph{problems}  (see Fig.~\ref{fig:mpe}).

To establish parallels between algorithms of diverse natures, we must first clearly answer: What is the essence of an MPE algorithm? And what do we mean, when we say that two algorithms are \emph{equivalent}?
The answers to these questions are not trivial, and crucial to the interpretation of the results of Secs.~\ref{subsec:sbm}--\ref{subsec:dc_sbm}.
Our goal with the next four subsections is therefore to clarify these issues.

\subsection{Anatomy of a black box}
\label{subsec:anatomy}

There are essentially two possible ways to formulate our answers, depending on how we think of MPE algorithms.

First, we may take the empirical point of view and declare that the essence of algorithms is their \emph{action}, independent of their inner workings.
According to this point of view, equivalence is functional and context dependent: If two algorithms give the same result on a series of networks $G_1,G_2,...G_k$, then the algorithms are equivalent with respect to these $k$ networks.
This allows us to treat algorithms as black boxes: Network in, partition out.
It is certainly an appealing approach, because it may be used to compare algorithms of widely different natures---say a genetic algorithm with an evolved objective function and a label propagation method.
Functional equivalence, however, has the drawback that it depends on the context, which makes it hard to draw definitive conclusions about algorithms.
Furthermore, it may identify somewhat artificial parallels, because it is insensitive to the \emph{origin} of the equivalences.

\begin{figure}
    \subfloat[]{
    \includegraphics[scale=0.5]{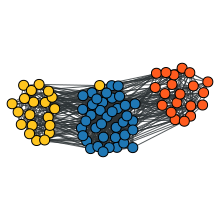}
    }
    \subfloat[]{
    \includegraphics[scale=0.5]{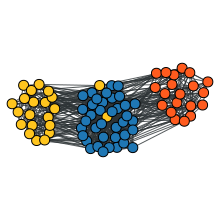}
    }
    \subfloat[]{
    \includegraphics[scale=0.5]{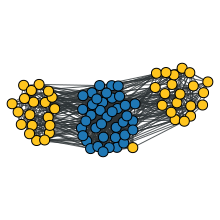}
    }
    \caption{Two--part algorithms in practice.
             We apply different combinations of maximizers and objective functions to a simple network with a clear block structure, generated using the SBM.
             Different MPE algorithms reveal different mesoscopic structures, but changing the objective function has the largest impact.
             (a, b) Same objective functions (modularity, see Sec.~\ref{subsubsec:modularity}) and different maximizers [(a) spectral, (b) greedy].
             (b, c) Same maximizer (greedy) with two different objective functions [(b) modularity, (c) core-periphery see Sec.~\ref{subsubsec:combinatorial}].
             The greedy maximizer is adapted from the Kernighan–Lin algorithm \cite{kernighan1970efficient}, in the spirit of Ref.~\cite{Karrer2011}, and the spectral maximizer relies on the embedding of the $q-1$ leading eigenvectors of the modularity matrix  \cite{Newman2004,Newman2006} in $\mathbb{R}^{q-1}$, followed by a clustering step, here implemented using a Gaussian mixture \cite{white2005spectral}. 
             }
    \label{fig:twopart}
\end{figure}

A second point of view is centered on the \emph{definitions} of MPE algorithms rather than their action therefore appears necessary.
Due to the diversity of existing MPE algorithms, this point of view will only be useful if we are able to first express MPE algorithms in some canonical form that can be readily analyzed.
One possibility is a two--part model expressed as the coupling of (i) an objective function that induces a total ordering of the partitions, and (ii) a maximizer that can find a---potentially local---optimum of the objective function  (see Fig.~\ref{fig:twopart}).
This two--part model captures the two important mechanisms that any MPE algorithm must possess.
On the one hand, the objective function captures the notion of quality of the partition and, consequently, tells the algorithm when to stop, and what partition to prefer whenever it has a choice.
On the other hand, the maximizer provides a mean of moving in the solution space, and of pinpointing the best partitions, as per the above criterion.
These mechanisms might be interwoven or hidden---we will touch on the subject shortly---, but the separation holds quite generally.

With the two--part model in place, equivalence takes on a crisp and clear meaning.
Two algorithms are either partially equivalent---same objective \textbf{or} same maximizer---or completely equivalent---same objective \textbf{and} same maximizer.
In the present paper, we will focus on partial equivalence, essentially ignoring  the maximizers.
This choice is motivated by the observations that (a) maximizers are, by necessity \footnote{ MPE problems are quite generally in \textsc{np-hard} \cite{Brandes2006,Crescenzi1995}.}, only efficient heuristics designed to find ``good enough'' optima in the rugged landscape of partitions \cite{Stein2013,Moore2017,Peel2017} (b) the no free lunch theorem implies that different objective functions and different inputs are associated with different optimal maximizers \cite{Peel2017}.
Hereafter, by \emph{equivalence}, we will therefore refer to the equivalence of the objective functions used.

\subsection{A glance under the hood}

In the simplest---and quite common---case, the separation in two parts is explicit.
For example, modularity--based methods famously attempt to maximize the modularity function
over the set  of all partitions of a network \cite{Fortunato2010}.
If there are many modularity optimization \emph{algorithms}, it is because there are many different mechanisms that can propose and refine partitions to find the optima of the modularity, e.g., the iterative spectral method of Ref.~\cite{Newman2006}, the fast unfolding method of Ref.~\cite{blondel2008fast}, or the message-passing algorithm of Ref.~\cite{Zhang2014b}.
The two--part algorithmic model is an exact description of these methods because they are framed in the language of objective functions.

Importantly, the two--part algorithmic model also holds in many cases where the emphasis is shifted away from an explicit objective function and maximizer dichotomy.
Consider as an example the classical label propagation algorithm of Ref.~\cite{raghavan2007near}.
This algorithm moves through the partition space by first assigning temporary labels (blocks) to nodes, and then repeatedly updating the labels with a majority rule (a node takes the label worn by  the majority of its neighbors).
Optimality is thus not defined for arbitrary pairs of partitions; it is instead expressed as a dynamical, initial condition dependent concept.
But a description in two parts can still be given, provided that we do some translation work:
The label propagation mechanism can be thought of as a maximizer, which naturally leads to \emph{partition flow} as a notion of optimality.
A partition $\mathcal{B}_1$ is better than $\mathcal{B}_2$ if the algorithm goes from $\mathcal{B}_2$ to $\mathcal{B}_1$ when it updates labels based on the majority rule.
With this definition, the best partitions are those that are stable against majority updates, and they are found via the propagation of labels.
One can construct an objective function with these orderings, and therefore a two--part algorithm indistinguishable from the original \cite{tibely2008equivalence,barber2009detecting,ver2014phase}.
\subsection{General graphical objective function}
Having established that a separation of algorithms into an objective function and a maximizing mechanism is often possible, let us turn to the functions themselves.

The outcome of pairwise interactions determines the structure of a complex network.
A \emph{general} objective function devised to uncover the mesoscopic patterns of a network therefore ought to include all these interactions in its calculation, at the very least.
If it does no more than that, then the function can be called \emph{graphical}, in the sense that no high-order terms are considered (i.e., there are no direct dependency on triplet of nodes, etc.).
From this point onward, we will focus on graphical objective functions alone;
the remainder of this paper is a testament to the generality of such a ``limited'' approach.

The definition of graphical objective function begins with the definition of its basic elements: Scores associated with each pair of nodes.
For the sake of generality, we will define these scores as real-valued functions, with essential dependencies on the partition $\mathcal{B}=\{B_1,\hdots,B_q\}$ under consideration, on the structure of the network as encoded by the  $n\times n$ adjacency matrix $\bm{A}$, and on an additional $n\times n$ side-information matrix $\bm{\lambda}$ that contains any pairwise information not directly captured by $\bm{A}$.
Let us therefore write the score associated with the pair of nodes $(i,j)\in[n]\times[n]$ (we use the integers $[k]=\{1,...,k\}$ to denote the nodes) as
\begin{equation}
    \label{eq:general_pair_score}
    f(a_{ij},\lambda_{ij}, \sigma_i,\sigma_j),
\end{equation}
where $\sigma_i\in[q]$ is the index of the block of node $i$, i.e., $\sigma_i = r$ if and only if $i\in B_r$.
We then express the aggregate of these local scores as
\begin{equation}
    \label{eq:general_objective_function}
    H\bigl(\bm{A},\bm{\lambda},\bm{\sigma};f\bigr) = \sum_{i,j: 1 \leq i \leq   j \leq n} f(a_{ij},\lambda_{ij}, \sigma_i,\sigma_j) \;,
\end{equation}
yielding a global objective function based on pairwise scores.
This form highlights the close parallel that exists between graphical objective functions and Edward--Anderson Hamiltonians \cite{Stein2013}, explicitly harnessed in a number of specific cases in Refs.~\cite{Reichardt2008,Reichardt2006,Decelle2011a,ronhovde2010local,kirkpatrick1983optimization}, for example.
The choice of a sum is, otherwise, for mere convenience; a product aggregate could have been equivalently implemented by taking $f\mapsto \log f$ and $H \mapsto e^H$.

\subsection{Equivalence and hierarchy under specialization}
\label{subsec:equiv_hierarhcy}
The last piece of the theoretical framework is a clear notion of connections among functions.
We use two concepts to establish these connections: Equivalence and specialization.

\subsubsection{Equivalence}
We say that two objective functions are equivalent if they induce the same total ordering of partitions, regardless of their inputs.
This definition captures the correct notion of equivalence, because it is clear that two equivalent objective functions---by this standard---will yield two MPE indistinguishable algorithms when they are  paired with the same maximizer.
As it stands, however, this notion of equivalence is not easy to handle mathematically.
We therefore resort to a second, stronger, criterion that leads to a more direct comparison procedure.
It is obvious that if
\begin{gather}
    H\bigl(\bm{A},\bm{\lambda},\bm{\sigma};f\bigr) < 
    H\bigl(\bm{A},\bm{\lambda},\bm{\sigma}';f\bigr)\implies
    g \circ H\bigl(\bm{A},\bm{\lambda},\bm{\sigma};f\bigr) < 
    g \circ H\bigl(\bm{A},\bm{\lambda},\bm{\sigma}';f\bigr)\;,
    \label{eq:strictly_increasing}
\end{gather}
for some strictly increasing function  $g$, then $H$ and $g \circ H$ are equivalent according to the first definition.
While this second version is more restrictive, it reduces the comparison of objective functions to the identification of the transformation $g$---an often straightforward process.

As we will see in Sec.~\ref{sec:hierarchy}, in practice, an even stronger criterion that limits $g$ to a particular subset of all  \emph{linear} transformation will often suffice to establish many equivalence relations.
Namely, whenever a pairwise score functions $f(a_{ij},\lambda_{ij}, \sigma_i,\sigma_j)$ can be split as
\begin{equation}
    \label{eq:split_condition}
    f(a_{ij},\lambda_{ij}, \sigma_i,\sigma_j)= f_1(a_{ij},\lambda_{ij}, \sigma_i,\sigma_j) +  f_2(a_{ij},\lambda_{ij}),
\end{equation}
where $f_2$ does not depend on the partition, we will be able to rewrite
the global objective function  as
\begin{align}
    H\bigl(\bm{A},\bm{\lambda},\bm{\sigma};f\bigr) &= \sum_{i\leq j}  f_1(a_{ij},\lambda_{ij}, \sigma_i,\sigma_j) + \sum_{i\leq j}  f_2(a_{ij},\lambda_{ij})\notag\\
    &\sim \sum_{i\leq j} f_1(a_{ij},\lambda_{ij}, \sigma_i,\sigma_j)\notag\\
    & = H'\bigl(\bm{A},\bm{\lambda},\bm{\sigma};f\bigr)
\end{align}
where ``$\sim$'' denotes equivalence, and ``$i\leq j $'' is a shorthand for the more precise statement ``$i,j: 1 \leq i \leq   j \leq n$.''
The equivalence holds because the additive terms are independent from $\bm{\sigma}$ and therefore do not affect the ordering.
Thus, equivalence will often follow from a simple linear transformation of the form $g \circ H=H - \sum_{i\leq j}  f_2(a_{ij},\lambda_{ij})$.

\subsubsection{Specialization}
With specialization, we aim to capture the idea that an objective function can be less expressive than its \emph{parent} function, i.e., that it is possible to fix some parameters of a function (the parent) to obtain a ``simpler'' version of the function \footnote{Formally, $f$ is in fact a placeholder for a function space with some parametrization set $\bm{\pi}$; $f$ only represent a unique function upon choosing some $p\in \bm{\pi}$. Specializing $f$ corresponds to defining a function space $f_S$ associated with a parameter set $\bm{\pi}_S\subseteq\bm{\pi}$.}.
It is more straightforward to define specialization at the level of pairwise score, and so we will say informally that a pairwise score function $f_S$ is a specialization of $f$ if $f_S$ is constructed by fixing some of the free parameters of $f$, in a way that alters the ranking of partitions, for some inputs.
Furthermore, we will say that the objective function $H'\bigl(\bm{A},\bm{\lambda},\bm{\sigma};f_S\bigr)$ is a specialization of the objective function $H\bigl(\bm{A},\bm{\lambda},\bm{\sigma};f\bigr)$ when $f_S$ is a specialization of $f$.

In the context of MPE, if $f_S$ is derived from $f$ and there exists at least one pair of nodes $(i,j)$ such that
\begin{gather}
    f_S(a_{ij},\lambda_{ij}, \sigma_i,\sigma_j)
    = 
    f_S(a_{ij},\lambda_{ij}, \sigma_i',\sigma_j')\notag\\
    \quad\text{and}\quad\notag\\
    f(a_{ij},\lambda_{ij}, \sigma_i,\sigma_j)
    \neq
    f(a_{ij},\lambda_{ij}, \sigma_i',\sigma_j')\;,
    \label{eq:specialization}
\end{gather}
where $\bm{\sigma}\neq \bm{\sigma}'$, then $f_S$ is a specialization of $f$ (and similarly for the resulting $H'$ and $H$).

Specialization is, in a sense, a one-way operation, because it involves reducing the complexity of a function.
In Eq.~\eqref{eq:specialization}, $f$ could act as $f_S$ but not the other way around, because $f_S$ is derived from $f$ by specialization.
Thus, specialization induces a hierarchy, with the most general functions at the top, and the most specialized ones at the bottom.
This is the hierarchy that we propose to delineate in the next sections.

\section{Objective function hierarchy under specialization}
\label{sec:hierarchy}

Recall that our claim is essentially the following: The objective functions of many mesoscopic pattern extraction algorithms are, in fact, special cases of the maximum likelihood formulation of the SBM.
Sections \ref{subsec:sbm} and \ref{subsec:dc_sbm} are devoted to showing how this comes about.
We begin with the methods that \emph{do not} account for any side information $\lambda_{ij}$, in Sec.~\ref{subsec:sbm}.
We show that they can be understood as specialization of the maximum likelihood formulation of the classical SBM \cite{Holland1983}.
We then move on to general MPE methods, in Sec.~\ref{subsec:dc_sbm}, by adding a side information dependency to the score functions.
Again, we show that these methods can be seen as specializations of a generalized SBM, close in spirit to the degree--corrected SBM of Ref.~\cite{Karrer2011}.
This part of the hierarchy sits above the methods of Sec.~\ref{subsec:sbm}, since the generalized SBM contains the classical SBM as a special case.
We summarize the relations between the various methods in  Fig.~\ref{fig:hierarchy}, and we show in Fig.~\ref{fig:examples} that they can be used to extract various patterns from a same real network.

\subsection{Partial hierarchy (no side information)}
\label{subsec:sbm}

\subsubsection{Stochastic block model}
\label{subsubsection:psbm}
Our starting point is the stochastic block model (SBM).
It is not an MPE algorithm \emph{per se}, but rather a random network model, amenable to statistical inference.
It prescribes a likelihood for the network $G$, parametrized by a latent partition $\mathcal{B}$ of its nodes.
The SBM becomes a MPE algorithm once this likelihood is used to infer the hidden partition $\mathcal{B}$ of $G$.
Although there are many ways of harnessing the likelihood to extract the mesoscopic patterns encoded by $\mathcal{B}$, we will only focus on \emph{likelihood maximization}, because it directly fits within the two--part model of MPE algorithms defined in Sec.~\ref{subsec:anatomy}; the likelihood is the objective function and the maximizer does not matter.

Given a network and a partition of the nodes in blocks associated with the vector $\bm{\sigma}$, the classical SBM \footnote{We base our derivation on the Poisson SBM, nearly equivalent to the somewhat more standard Bernoulli SBM. Our choice is justified by the fact that the full hierarchy of objective functions follows more naturally from the Poisson SBM.}
prescribes that the number of edges between nodes $(i,j)$ should be drawn from a Poisson distribution of mean $\omega_{\sigma_i\sigma_j}$.
All edges are assumed to be independent, such that the likelihood of the complete graph is given by
\begin{equation}
    \label{eq:sbm_micro_likelihood}
    \mathbb{P}(G|\mathcal{B},\bm{\omega}) = \prod_{i \leq j} \frac{(\omega_{\sigma_i\sigma_j})^{a_{ij}}}{a_{ij}!}e^{-\omega_{\sigma_i\sigma_j}}\;.
\end{equation}
It is parametrized by the $q\times q$ matrix $\bm{\omega}$ and the partition $\mathcal{B}$ (or equivalently by the block assignments $\bm{\sigma}$).
The standard inference procedure calls for the estimation of both, $\bm{\omega}$ and $\mathcal{B}$, usually through alternated learning of the two sets of parameters (via the expectation--maximization algorithm \cite{Decelle2011a}).
However, we will focus on the estimation of $\mathcal{B}$ alone, treating the parameters $\bm{\omega}$ as ``control buttons.''
The freedom to \emph{impose} parameters $\bm{\omega}$ on the network will ultimately allow us to draw relations with other MPE algorithms.

To extract $\mathcal{B}^*(G)$---the ``true'' partition of the nodes---from the network, we maximize the likelihood of the SBM  with respect to the partition (see also Sec.~\ref{sec:discuss}).
Since the logarithm is a strictly increasing function of its argument, we may equivalently maximize the log-likelihood
\begin{equation}
    \label{eq:loglikelihood_sbm}
    \log \mathbb{P}(G|\mathcal{B},\bm{\omega}) = \sum_{i \leq j}\left[a_{ij}\log \omega_{\sigma_i\sigma_j} - \omega_{\sigma_i\sigma_j} - \log a_{ij}!\right]\;.
\end{equation}
This is a first (trivial) example of the concept of equivalence of Sec.~\ref{subsec:equiv_hierarhcy}.
It becomes evident upon inspection of Eq.~\eqref{eq:loglikelihood_sbm} that the log-likelihood is, in fact, a graphical objective function of the general form appearing in Eq.~\eqref{eq:general_objective_function}, associated with the pairwise score function
\begin{equation}
    \label{eq:sbm_pairwise_score}
    f_{\mathrm{SBM}}(a_{ij}, \sigma_i,\sigma_j) \sim a_{ij}\log \omega_{\sigma_i\sigma_j} - \omega_{\sigma_i\sigma_j} \;.
\end{equation}
Thus, any objective function that can be written as a special case of Eq.~\eqref{eq:sbm_pairwise_score} will be a specialization of the maximum likelihood formulation of the SBM.

\subsubsection{General modular graph model}
\label{subsubsec:gmgm}
One such (explicit) specialization is the general modular graph model (GMGM) \cite{Kawamoto2017,Young2017}.
Like its general counterpart, the GMGM is a generative model for networks that supposes a latent partition of the nodes in blocks.
The crucial difference is that the connection matrices $\bm{\omega}$ of the GMGM are much more structured than that of the SBM.
  
Pairs of blocks are assigned one of two types, say $a$ and $b$, and this information is encoded in a $q\times q$ binary (and symmetric) matrix $\bm{X}$.
If a pair of blocks ($B_r, B_s$) is of type $a$, then we set $x_{rs}=1$.
Contrariwise, we set $x_{rs}=0$ if the pair $(B_r,B_s)$ is of type $b$.
Pairs of blocks of type $a$ are then \emph{all} associated with a connectivity $\omega_{rs}$ = $\omega_{a}$, while pairs of  type $b$ are associated with a connectivity $\omega_{rs}=\omega_{b}$, where 
we take $\omega_b<\omega_a$ without loss of generality \footnote{%
A parametrization where $\omega_b>\omega_a$ can be represented by an equivalent parametrization $(\bm{X}',\bm{\omega}')$, defined as $\bm{X}'=\bm{11}^T-\bm{X}$ with $\omega_a'=\omega_b$, and $\omega_b'=\omega_a$.
The case $\omega_{a}=\omega_{b}$ is somewhat pathological, but it can be handled nonetheless, with $\bm{X}=\bm{11}^T$, $\omega_{a}'=\omega_{a}$ and $\omega_{b}'=\delta$, for any  $\delta<\omega_{a}$.}.
Every connection matrix of the GMGM can therefore be written as
\begin{equation}
    \label{eq:matrix_gmgm}
    \bm{\omega} = \omega_{b}\bm{11}^\T+ (\omega_{a}-\omega_{b}) \bm{X}
\end{equation}
where $\bm{1}$ is column vector of ones.

The principal motivation for using the simplified matrices of Eq.~\eqref{eq:matrix_gmgm} is that the mathematical treatment of the model becomes simpler at the expense of a moderately reduced flexibility \cite{Kawamoto2017,Young2017}.
In particular, the two identities (used in similar derivations in Refs.~\cite{newman2016equivalence,newman2013community})
\begin{subequations}
\label{eq:gmgm_trick}
\begin{align}
    \omega_{rs} &=  \omega_{b} + x_{rs}(\omega_{a} - \omega_{b}) \;,\\
    \log \omega_{rs} &= \log \omega_{b} + x_{rs} (\log \omega_{a} -  \log \omega_{b})\;,
\end{align}
\end{subequations}
lead to a likelihood and a log-likelihood analogous to---but much simpler than---the ones appearing in Eqs.~\eqref{eq:sbm_micro_likelihood} and \eqref{eq:loglikelihood_sbm}.
They are associated with the score function 
\begin{align}
    f_{\mathrm{GMGM}}(a_{ij}, \sigma_i,\sigma_j)
     &= a_{ij}\left[\log\omega_b + x_{\sigma_i\sigma_j}(\log\omega_a - \log\omega_b) \right]- \left[\omega_b +  x_{\sigma_i\sigma_j}(\omega_a - \omega_b)\right] \notag\\
    &\sim x_{\sigma_i\sigma_j}[a_{ij} (\log\omega_a - \log\omega_b) - (\omega_a - \omega_b) ]\notag\\
    &\sim x_{\sigma_i \sigma_j}\big[a_{ij} + \gamma\big]
    \label{eq:gmgm_pairwise}
\end{align}
where $\gamma= (\omega_b - \omega_a) / (\log \omega_a -\log \omega_b) \in(-\infty,0]$, a drastic simplification when contrasted with Eq.~\eqref{eq:sbm_pairwise_score}.
In essence, the GMGM only cares about the type of a block pair.
If a pair of nodes $(i,j)$ is associated with a block pair of type $a$, then the global objective function is increased by a factor of $a_{ij} + \gamma$ (greater when $a_{ij}=1$ than when $a_{ij}=0$).
If a pair of nodes $(i,j)$ is associated with a block pair of type $b$, then it only has an indirect impact, by omission.

\subsubsection{Combinatorial objective functions}
\label{subsubsec:combinatorial}
The GMGM specialization of the SBM is interesting not only for its mathematical simplicity, but also  because its pairwise score function can be obtained from a completely different perspective.
As we have seen, the essence of the GMGM is its binary classification of block pairs; it turns out that there are countless examples of MPE objective functions that rely on a similar dichotomy (see, for instance, Ref.~\cite{Fortunato2016} for a recent review).
Their design is essentially the following.
Some subsets or intersections of nodes are  first identified as special.
The MPE objective function is then designed as to maximize the number of edges \emph{within} or \emph{to} these subsets.
Finally, because there are often trivial maxima (e.g., place all nodes in the special subset), some constraints are added to avoid trivial optima.

The general mathematical construction closely parallels that of the GMGM.
First, we designate special pairs of blocks, and encode the result in a binary matrix $\bm{X}$.
We then assume, without loss of generality, that the number of edges within these blocks should be \emph{maximized} by the target partition $\mathcal{B}^*$.
This leads to the graphical objective function
\begin{equation}
    \label{eq:unconstrained_obj_func}
    \widetilde{H}(G|\mathcal{B}) = \sum_{i\leq j} a_{ij} x_{\sigma_i\sigma_j}\;.
\end{equation}

Functions of the form of Eq.~\eqref{eq:unconstrained_obj_func}  are plagued by many trivial optima, since it is often possible to maximize $\widetilde{H}$ by placing all nodes in one or a few blocks.
For instance, if $x_{rr}=1$ for at least one $r$, then Eq.~\eqref{eq:unconstrained_obj_func} is maximized by putting all nodes in block $B_r$---it is obvious that no mesoscopic information is contained in the resulting partition.
In general, if Eq.~\eqref{eq:unconstrained_obj_func} rewards placing many edges between some pair of blocks $(B_r,B_s)$ via $x_{rs}=1$,  then it is possible to find good solutions simply by putting a lot of \emph{nodes} in these blocks: The more nodes, the more edges, and therefore the better score.
We discourage these uninformative solutions by introducing an additive \emph{balance constraints} $h(\mathcal{B})$ that penalizes the objective function $\widetilde{H}$ for partitions that contain large blocks aligned with $\bm{X}$.
Specifically, we use a quadratic constraint on the block sizes \cite{sulc2010belief,newman2013community}
\begin{equation}
    \label{eq:bal_constraint_def}
   h(\mathcal{B}) = 2\gamma \sum_{r,s} x_{rs}n_rn_s\;, \qquad \gamma<0\;,
\end{equation}
where $n_r$ is the size of block $B_r$, and where $|\gamma|$ controls the overall strength of the constraint $h$.
Because the constraint appearing in Eq.~\eqref{eq:bal_constraint_def} can be rewritten as
\begin{align*}
 2\gamma\sum_{r, s} x_{rs}n_rn_s &= 2 \gamma\sum_{r, s} x_{rs} \left(\sum_{i=1}^n \delta_{\sigma_i r}\right)\left(\sum_{j=1}^n \delta_{\sigma_j s}\right) \\&= \gamma\sum_{i\leq j} x_{\sigma_i\sigma_j}
\end{align*}
where $\delta_{ab}$ is the Kronecker delta (equal to $1$ if $a=b$ and to zero otherwise), the constrained version of Eq.~\eqref{eq:unconstrained_obj_func} is equivalent to
\begin{align}
    \label{eq:counts_plus_quad_bal}
    H(G|\mathcal{B}) = \widetilde{H}(G|\mathcal{B}) + h(\mathcal{B}) = \sum_{i\leq j} x_{\sigma_i \sigma_j}[ a_{ij} + \gamma]\;.
\end{align}
This balanced objective function is obviously associated with a pairwise score function equivalent to that of the GMGM [c.f. Eq.~\eqref{eq:gmgm_pairwise}].
Therefore, all objective functions formulated as an \emph{edge count maximization} coupled with an additive \emph{quadratic balance constraint} are equivalent to the GMGM.
Furthermore, the strength of the balance constraint $\gamma$ can be seen as a function of the parameters $(\omega_a,\omega_b)$ of the corresponding GMGM: The greater the difference between $\omega_a$ and $\omega_b$, the stronger the balance constraint.

The equivalence of the GMGM with combinatorial objective function has far reaching consequences, because many MPE methods are based on variation on these functions.
A few well-known examples are: Balanced minimum cut, with $\bm{X}=\bm{I}$ where $\bm{I}$ is an identity matrix \cite{Newman2013}; approximative graph coloring with $\bm{X}=\bm{11}^\T-\bm{I}$ \cite{krzakala2009hiding,Decelle2011a}; nonoverlapping core-peripheries (CP) under size constraints \cite{Rombach2014,kojaku2017finding} with, e.g.,
\begin{gather*}
    \bm{X}_{\mathrm{CP1}} = \begin{pmatrix} 1 & 0 \\ 0 & 0 \\ \end{pmatrix}\;,\qquad
    \bm{X}_{\mathrm{CP2}} = \begin{pmatrix} 1 & 1 \\ 1 & 0 \\ \end{pmatrix}\;,\qquad
    \bm{X}_{\mathrm{MultiCP}} =
    \begin{pmatrix}
    1 & 1 & \hdots & 0 & 0\\
    1 & 0 & \hdots & 0 & 0\\
     \vdots & \vdots& \ddots & \vdots & \vdots\\
    0 & 0 & \hdots & 1 & 1\\
     0 & 0 & \hdots & 1 & 0\\
    \end{pmatrix}\;.
\end{gather*}
If anything, these simple examples show that the GMGM and Eq.~\eqref{eq:counts_plus_quad_bal} can be used as an ``objective function  factory'' of sort:
For any choice of $\gamma$ and $q$, there will be $2^{\binom{q}{2}+q}$ different binary symmetric matrices $\bm{X}$, and as many MPE objective functions.
Those that are named and well studied are but a tiny fraction of the full spectrum of possibilities; most will uncover exotic patterns that are mixtures of core-peripheries, cuts, coloring, hierarchies, etc.

\begin{figure*}
    \centering
    \includegraphics{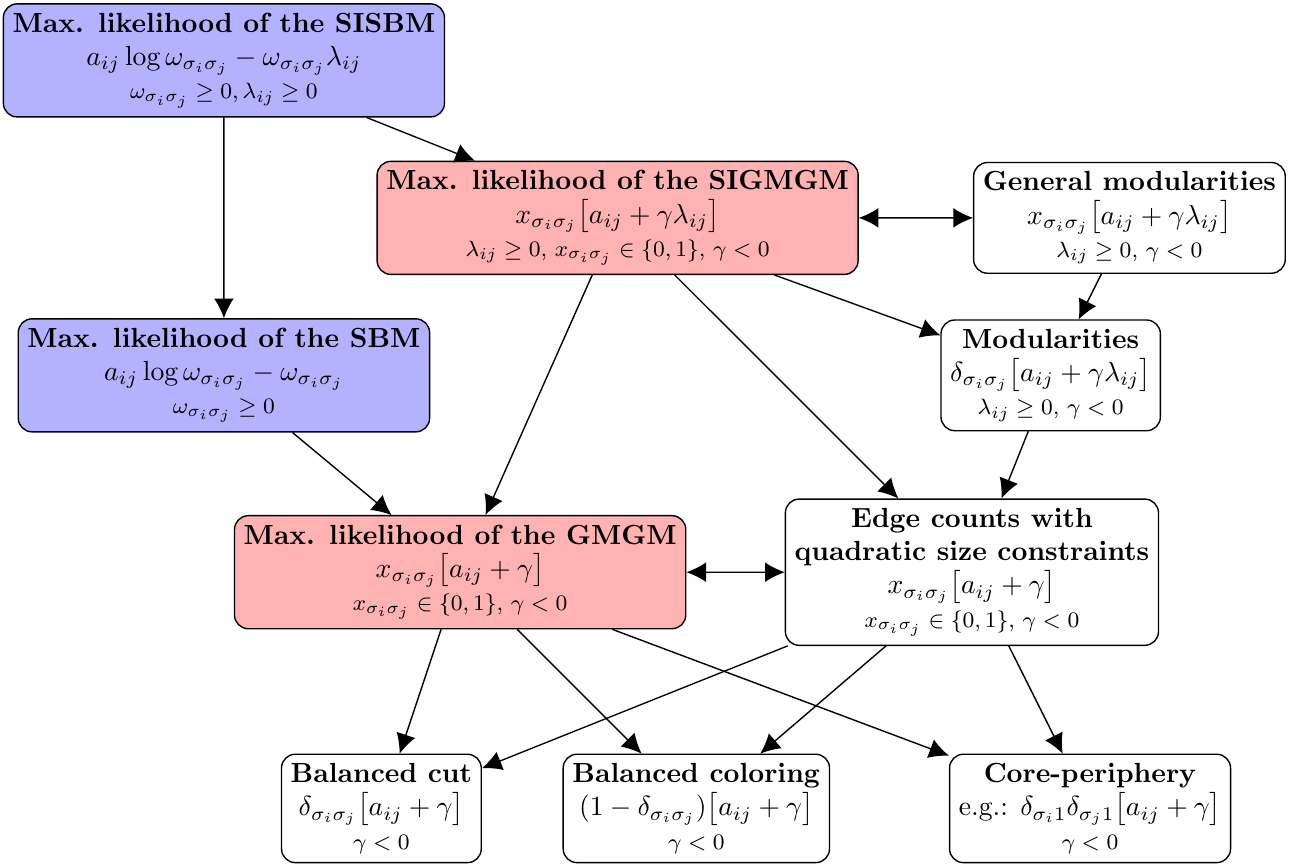}
    \caption{
    Partial hierarchy of objective functions.
    The pairwise score function of MPE methods are shown with the range of parameters below.
    Arrows denote specialization; doubled--sided arrows denote equivalence.
    Only the most direct arrows are drawn for the sake of clarity; specialization and equivalence are transitive operations.
    The abbreviations are: stochastic block model (SBM), with side information (SISBM); and
    general modular graph modular model (GMGM), with side information (SIGMGM).
    Functions derived from the perspective of statistical inference are colored in blue (general classification) and red (binary classification).
    }
    \label{fig:hierarchy}
\end{figure*}
\subsection{Complete hierarchy}
\label{subsec:dc_sbm}
While the SBM and its GMGM are general enough to specialize to many well-known MPE methods, there are also numerous objective functions that cannot be written as in Eqs.~\eqref{eq:sbm_pairwise_score} and \eqref{eq:gmgm_pairwise}---e.g., modularity functions---, because they rely also on some \emph{side-information} matrix $\bm{\lambda}$ absent from the pairwise scores of Eq.~\eqref{eq:sbm_pairwise_score}.
The purpose of the present section is to expand on the classification of Sec.~\ref{subsec:sbm} to accommodate these functions.

\subsubsection{Stochastic block model with side information}
\label{subsubsec:si_sbm}
In the spirit of Ref.~\cite{Karrer2011}, we define a generalization of the Poisson SBM, whose likelihood is given by
\begin{equation}
    \label{eq:sisbm_likelihood}
    \mathbb{P}(G|\mathcal{B},\bm{\omega},\bm{\Lambda}) = \prod_{i\leq j} \frac{(\omega_{\sigma_i\sigma_j}\lambda_{ij})^{a_{ij}}}{a_{ij}!}e^{-\omega_{\sigma_i\sigma_j}\lambda_{ij}}\;.
\end{equation}
This (over-parametrized) version of the SBM combines mesoscopic information (via $\bm{\omega}$) with side information at the level of edges (via $\bm{\Lambda}$).
It directly specializes to many well--known likelihoods, including the classical Poisson SBM (with $\bm{\Lambda}=\bm{11}^\T$), or the degree-corrected SBM of Ref.~\cite{Karrer2011} (with $\bm{\Lambda}=\bm{kk}^\T/2m$ where $\bm{k}$ is the vector of degrees).

As with its classical counterpart, one can find the most likely partition of the nodes of $G$  by maximizing the logarithm of the likelihood \eqref{eq:sisbm_likelihood}:
\begin{equation*}
    \log\mathbb{P} = \sum_{i\leq j}\left[ a_{ij}\log \omega_{\sigma_i\sigma_j}\lambda_{ij}  -\omega_{\sigma_i\sigma_j}\lambda_{ij} - \log a_{ij}!\right]\;.
\end{equation*}
Therefore, the maximum likelihood formulation of the  SBM with side information (hereafter: SISBM) is associated with the pairwise score function
\begin{equation}
    f_{\mathrm{SISBM}} (a_{ij},\lambda_{ij},\sigma_i,\sigma_j) \sim a_{ij}\log \omega_{\sigma_i\sigma_j}  -\omega_{\sigma_i\sigma_j}\lambda_{ij}\;.
    \label{eq:sisbm_pairwise_score}
\end{equation}
The likelihood is not useful in itself, because there are too many parameters for the amount of information encoded in $\bm{A}$.
However, considering Eq.~\eqref{eq:sisbm_pairwise_score} not as a proper MPE method, but rather as the starting point of a general objective function hierarchy,  it becomes a useful classification tool.

\subsubsection{General modular graph model with side information}
As with the classical SBM, it is possible to define a GMGM specialization of the SISBM.
Following Sec.~\ref{subsubsec:gmgm}, the idea is again to classify all pairs of blocks according to their density category (via $\bm{X}$), and to re-use the identities appearing in Eq.~\eqref{eq:gmgm_trick} to rewrite the log-likelihood.
The resulting likelihood is still over-parametrized because of $\bm{\Lambda}$, but much simpler than that of the general SISBM, since the connection matrices $\bm{\omega}$ are now restricted to the form of Eq.~\eqref{eq:matrix_gmgm}.
It is easy to show that the pairwise score function is now
\begin{align}
    f_{\mathrm{SIGMGM}}(a_{ij},\lambda_{ij},\sigma_i,\sigma_j)
     &= a_{ij}\big[\log \omega_{b} + x_{rs} (\log \omega_{a} -  \log \omega_{b})\big]- \big[ \omega_{b} + x_{\sigma_i\sigma_j}(\omega_{a} - \omega_{b})\big]\lambda_{ij}\notag\\
    &\sim  x_{\sigma_i\sigma_j}\big[ a_{ij} (\log \omega_{a} -  \log \omega_{b}) - \lambda_{ij}(\omega_{a} - \omega_{b})\big]\notag\\
    &\sim  x_{\sigma_i\sigma_j}\big[ a_{ij} +\gamma \lambda_{ij} \big]
    \label{eq:sigmgm_pairwise}
\end{align}
where $\gamma<0$ is the same parameter as the one appearing in Eq.~\eqref{eq:gmgm_pairwise}.

\subsubsection{Modularity functions}
\label{subsubsec:modularity}
One of the reasons why the GMGM specialization is useful is, again, that it can be derived from first principles in a completely different manner, this time from the point of view of the \emph{modularity} \cite{Newman2004}.
In a nutshell, modularity is defined as the difference between the number of internal edges of a partition (edges that connect two nodes in the same block), and the \emph{expected} number of internal edges for this partition, if the network were to be drawn from some null model.
The idea behind modularity is to maximize the number of edges within blocks, while accounting for the edges that would have been there in the first place, just by pure chance (assuming some model for the network, see Table.~\ref{table:null_models}). 

\begin{table}
    \begin{ruledtabular}
        \begin{tabular}{lcl}
            Model & $\lambda_{ij}$ &  Ref.\\
            \hline
        Configuration model (CM) & $k_ik_j/(2m)$&  \cite{Newman2004}\\
        CM with resolution & $\zeta k_ik_j/(2m)$&  \cite{Reichardt2006}\\
        Erd\H{o}s-R\'enyi & $\rho$ & \cite{Traag2011}\\
        Constant Potts model &$ \zeta$ & \cite{Traag2011}\\
        Gravity model&  $k_ik_j\phi(r_{ij})$ &  \cite{expert2011uncovering}\\
        \end{tabular}
    \end{ruledtabular}
    \caption{Examples of null models. In all cases, $\zeta>0$ is a free parameter, $\rho\in[0,1]$ is the density of the network. The gravity model is included as an example of an exotic null model; it is derived for spatially embedded network, with $r_{ij}$ being the Euclidean distance between nodes $i$ and $j$  and $\phi$ some reference connection propensity in space.
    \label{table:null_models}
    }
\end{table}

Modularity is a graphical objective function, since it can be written as a sum over pairs of nodes \cite{Newman2006,Reichardt2006}.
Writing the expected number of edges between the nodes $i$ and $j$ as $\lambda_{ij}$ under the null model of choice, the modularity of a partition reads
\begin{equation}
    \label{eq:modularity}
    H_{\mathrm{Mod}}(\mathcal{B},\bm{\lambda}, G) \propto \sum_{i\leq j} [a_{ij}-\lambda_{ij}]\delta_{\sigma_i\sigma_j}\;.
\end{equation}
Importantly, the pairwise score function associated with the modularity is always given by
\begin{align}     
    f_{\mathrm{Mod}}(a_{ij},\lambda_{ij},\sigma_i,\sigma_j) &= (a_{ij} -\lambda_{ij})\delta_{\sigma_i\sigma_j} \sim (a_{ij} +\gamma \widetilde{\lambda}_{ij})\delta_{\sigma_i\sigma_j} \;,
    \label{eq:score_modularity}
\end{align}
independent from the choice of null model (where $\gamma<0$ and $\widetilde{\lambda}_{ij}$ is a rescaled connection probability under the null model).

A comparison with Eq.~\eqref{eq:sigmgm_pairwise} reveals that the above score function---and therefore \emph{any} modularity-type function---is in fact a  specialization of the GMGM with side information, recovered by setting $\bm{X}=\bm{I}$, $\gamma=-1$, and by using the null model of the modularity as the side-information matrix $\bm{\Lambda}$.
In other words, every modularity function is equivalent to some variant of the SIGMGM, where the null model is multiplied with the connection matrix $\bm{\omega}$ where $\bm{X}$ is simply the identity.
\begin{figure*}
    \centering
    \includegraphics{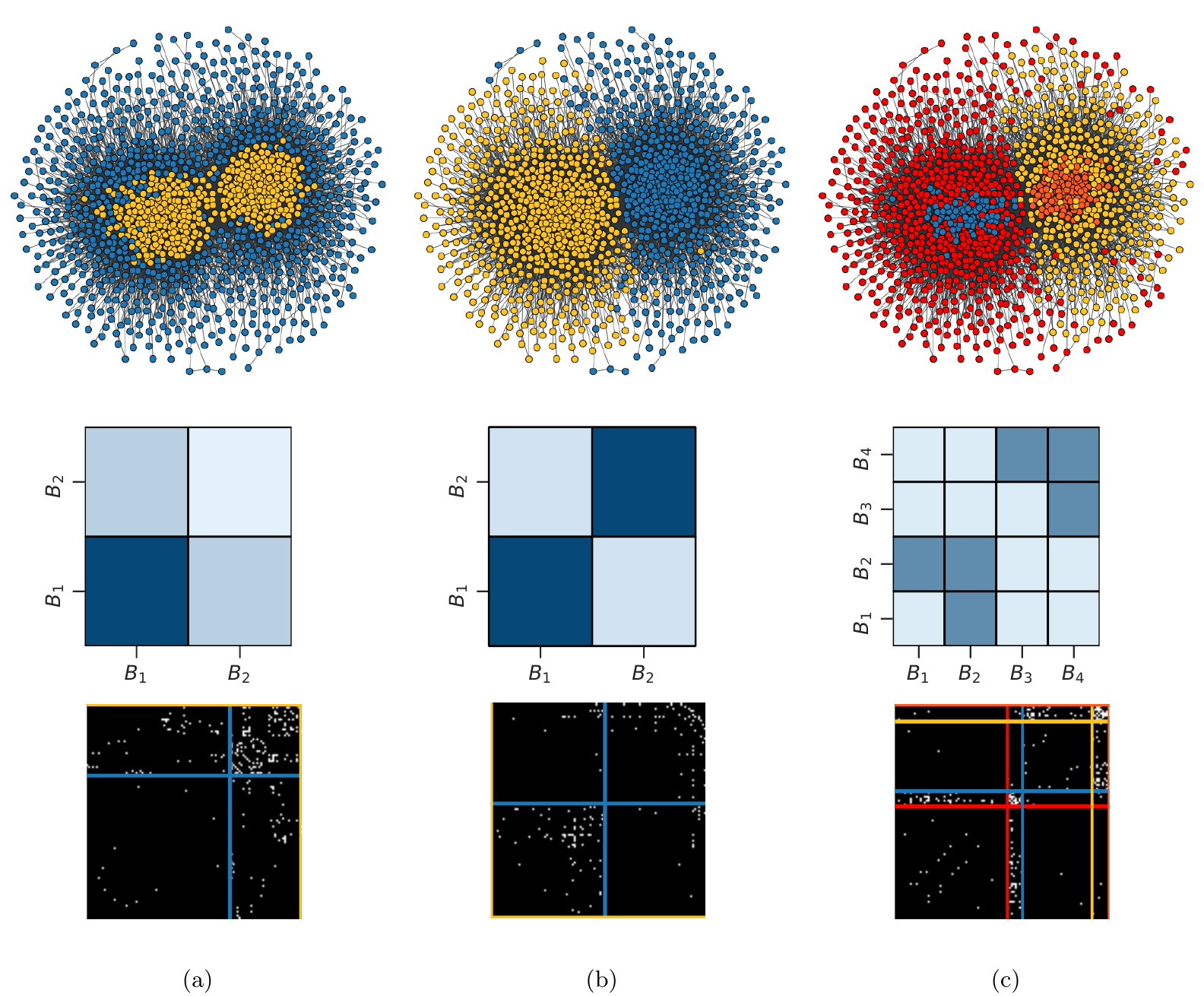}
    \caption{
        Mesoscopic patterns learned from and imposed on a real complex network.
        All results are obtained on the \texttt{polblog} dataset, a directed network of hyperlinks between weblogs on US politics, recorded shortly after the 2004 presidential election. 
        There are a total of $1\;222$ nodes (weblogs) and $16\;714$ edges.
        We use an undirected, self-loop free version of the network.
        All subfigures show (top) the network with nodes colored according to the identified partition, (center) a cartoon of the matrix $\bm{\omega}$ \emph{imposed} for the equivalent SBM [darker shades of blue represent larger values of $\omega_{rs}$], and (bottom) the adjacency matrix with the limit of blocks indicated as colored lines and edges as white dots.
        The optima of the objective functions are found via simulated annealing and greedy search \cite{Young2017}.
        (a) Natural partition of the network in $q=2$ blocks, as found with the classical Bernoulli SBM via expectation--maximization (EM) on $\mathcal{B}$ and $\bm{\omega}$. In this case alone, $\bm{\omega}$ is learned and not imposed.
        (b) Balanced cut obtained with $\gamma\approx-25$ and the GMGM. The two blocks have size $\bm{n}^\T=[650, 572]$. A similar partition is identified by the modularity.
        (c) Double core-periphery  found with $\gamma\approx-9$. The cores are of sizes 98 and 87 while their respective peripheries contain 396 and 641 nodes. There are only $3\;703$ edges between nodes of the peripheries (out of a maximum of $283\;330$ possible edges).
    }
    \label{fig:examples}
\end{figure*}

It is worth pointing out that using a flat null model (i.e., $\widetilde{\lambda}_{ij}=1$) in Eq.~\eqref{eq:score_modularity} amounts to opting for a GMGM without side information, with $\bm{X}=\bm{I}$. 
Because the latter is associated with a pairwise score functions that is equivalent to edge counts coupled with quadratic balance constraints, it follows that \emph{flat null models} act exactly like quadratic balance constraints.
This correspondence explains the regularization properties of the ER null model investigated in Ref.~\cite{Traag2011,ronhovde2010local} (among others).

We note in closing that while the connection between the modularity and the GMGM is presented here for $\bm{X}=\bm{I}$, it is of course possible to define ``modularities'' associated with different matrices $\bm{X}$, in the spirit of the side information free equivalence.
These modularities will be able to uncover any mixture of mesoscopic patterns reflected in $\bm{X}$.

\section{Discussion}
\label{sec:discuss}

In this paper, we have shown that the maximum likelihood formulation of the SBM is perfectly equivalent to a number of standard mesoscopic pattern extraction (MPE) methods, upon appropriate specialization of its density matrix $\bm{\omega}$.
Specifically, we have found that different classes of density matrices are associated with various classes of MPE algorithms, such as minimum cuts, modularities, core-periphery algorithm and combinations thereof.
This has allowed us to delineate a hierarchy of MPE methods (Fig.~\ref{fig:hierarchy}), and to understand all methods as increasingly simplified SBMs.
In doing so, we have shown that the SBM is universal with respect to mesoscopic pattern extraction with graphical functions---a conclusion that is complementary to the recent observation that the SBM is a universal network approximator \cite{olhede2014network}.

Apart from a better understanding of MPE methods, in the light of the hierarchy of Fig.~\ref{fig:hierarchy}, there are a number of practical consequences to the fact that many of the MPE methods of network science are, after all, the SBM in disguise.
Let us mention a few in closing.

First and foremost, these equivalences imply that the efficient maximizers (see, e.g., Refs.~\cite{Peixoto2014}) developed to tackle the hard problem of estimating $\mathcal{B}$ for the general SBM can be reused to solve more specific MPE problems that are also hard.
This application of the equivalences is direct: To optimize an MPE objective function, simply fix the matrix $\bm{\omega}$ (and $\bm{\Lambda}$ if there is side information) with some target mesoscopic pattern in mind, and run an SBM likelihood maximization  procedure to uncover $\mathcal{B}$ (we have used this method to obtain the results of Fig.~\ref{fig:examples}).

Second, as is also pointed out in Ref.~\cite{newman2016equivalence} (for the special case of modularity), arbitrary MPE methods that are specializations of the SBM now stand on sounder statistical foundations, once their connection with the SBM is recognized.
This is due to the fact that their free parameters---e.g. $\gamma$ in Eqs.~\eqref{eq:gmgm_pairwise} and \eqref{eq:sigmgm_pairwise}---can be interpreted as functions of the connectivity matrix $\bm{\omega}$, thereby providing a statistically principled estimation procedure---expectation--maximization \cite{Decelle2011a}---for otherwise arbitrary parameters.

Third, we can conclude that a number of hidden assumptions are built into popular MPE methods.
In particular, they amount to fitting simplified SBMs by maximum likelihood, often with misspecified density matrices $\bm{\omega}$.
Doing so is not a problem \textit{per se}, because the goal of MPE is not always to find the most natural or most statistically robust decomposition of a network \cite{Schaub2017}; it might instead be to of reveal  different facets of the mesoscopic organization of a same network (see Fig.~\ref{fig:examples}).
However, one should bear in mind that using these MPE methods amounts to fitting an ill-defined model, with all the problems that this may bring about, such as missing the best description of a network, or preventing inference algorithms from converging at all  \cite{kawamoto2017algorithmic,kawamoto2018algorithmic,Zhang2014b}.

Fourth, a knowledge of equivalences can help us better interpret the empirical outcomes of mesoscopic pattern extraction.
Two algorithms may behave similarly on a set of networks not due to the robustness of the patterns therein, but because they share an equivalent notion of optimality.
Hence, empirical studies that rely on many MPE algorithms---say, comparative analyses \cite{ghasemian2018evaluating,kawamoto2018comparative}---can avoid being lured by what appears to be a strong consensus of many methods that actually implement the same notion of optimality.

Finally, the equivalences lead to a number of theoretical shortcuts.
One, the consistency results derived for the SBM \cite{abbe2018community} apply directly to all MPE algorithms in the hierarchy, by specialization.
The consistency of the SBM in most scaling regimes (and the existence of a \emph{detectability limit} \cite{Decelle2011a}) therefore extends to virtually every MPE algorithm studied thus far.
Two, formal \textsc{NP}--hardness results can be extended to many MPE methods, using trivial reductions.
For example, since it is known that modularity maximization is in  \textsc{NP}--hard \cite{Brandes2006}, the equivalence of modularity with the likelihood maximization of the GMGM specialization of the degree-corrected SBM \cite{newman2016equivalence} directly implies the NP--hardness of the latter, and therefore of the SBM.
Three, the universality of the SBM suggests that there is an extension of the no free lunch Theorem of Ref.~\cite{Peel2017} to a more generalized notion of MPE problems---not just community detection.

\section*{Acknowledgments}
We thank Daniel Larremore, Edward Laurence, Charles Murphy and Laurent H\'ebert-Dufresne for useful comments.
This work was funded by the Fonds de recherche du Qu\'ebec-Nature et technologies (J.-G.Y., P.D.), the Natural Sciences and Engineering Research Council of Canada (G.S.O., L.J.D.), and the  Sentinel North program, financed by the Canada First Research Excellence Fund (J.-G.Y., G.S.O., P.D., L.J.D.).


\begin{thebibliography}{57}%
\makeatletter
\providecommand \@ifxundefined [1]{%
 \@ifx{#1\undefined}
}%
\providecommand \@ifnum [1]{%
 \ifnum #1\expandafter \@firstoftwo
 \else \expandafter \@secondoftwo
 \fi
}%
\providecommand \@ifx [1]{%
 \ifx #1\expandafter \@firstoftwo
 \else \expandafter \@secondoftwo
 \fi
}%
\providecommand \natexlab [1]{#1}%
\providecommand \enquote  [1]{``#1''}%
\providecommand \bibnamefont  [1]{#1}%
\providecommand \bibfnamefont [1]{#1}%
\providecommand \citenamefont [1]{#1}%
\providecommand \href@noop [0]{\@secondoftwo}%
\providecommand \href [0]{\begingroup \@sanitize@url \@href}%
\providecommand \@href[1]{\@@startlink{#1}\@@href}%
\providecommand \@@href[1]{\endgroup#1\@@endlink}%
\providecommand \@sanitize@url [0]{\catcode `\\12\catcode `\$12\catcode
  `\&12\catcode `\#12\catcode `\^12\catcode `\_12\catcode `\%12\relax}%
\providecommand \@@startlink[1]{}%
\providecommand \@@endlink[0]{}%
\providecommand \url  [0]{\begingroup\@sanitize@url \@url }%
\providecommand \@url [1]{\endgroup\@href {#1}{\urlprefix }}%
\providecommand \urlprefix  [0]{URL }%
\providecommand \Eprint [0]{\href }%
\providecommand \doibase [0]{http://dx.doi.org/}%
\providecommand \selectlanguage [0]{\@gobble}%
\providecommand \bibinfo  [0]{\@secondoftwo}%
\providecommand \bibfield  [0]{\@secondoftwo}%
\providecommand \translation [1]{[#1]}%
\providecommand \BibitemOpen [0]{}%
\providecommand \bibitemStop [0]{}%
\providecommand \bibitemNoStop [0]{.\EOS\space}%
\providecommand \EOS [0]{\spacefactor3000\relax}%
\providecommand \BibitemShut  [1]{\csname bibitem#1\endcsname}%
\let\auto@bib@innerbib\@empty
\bibitem [{\citenamefont {Moore}(2017)}]{Moore2017}%
  \BibitemOpen
  \bibfield  {author} {\bibinfo {author} {\bibfnamefont {C.}~\bibnamefont
  {Moore}},\ }\href {https://arxiv.org/abs/1702.00467} {\bibfield  {journal}
  {\bibinfo  {journal} {arXiv:1702.00467}\ } (\bibinfo {year}
  {2017})}\BibitemShut {NoStop}%
\bibitem [{\citenamefont {Fortunato}\ and\ \citenamefont
  {Hric}(2016)}]{Fortunato2016}%
  \BibitemOpen
  \bibfield  {author} {\bibinfo {author} {\bibfnamefont {S.}~\bibnamefont
  {Fortunato}}\ and\ \bibinfo {author} {\bibfnamefont {D.}~\bibnamefont
  {Hric}},\ }\href {\doibase 10.1016/j.physrep.2016.09.002} {\bibfield
  {journal} {\bibinfo  {journal} {Phys. Rep.}\ }\textbf {\bibinfo {volume}
  {659}},\ \bibinfo {pages} {1} (\bibinfo {year} {2016})}\BibitemShut {NoStop}%
\bibitem [{\citenamefont {Schaub}\ \emph {et~al.}(2017)\citenamefont {Schaub},
  \citenamefont {Delvenne}, \citenamefont {Rosvall},\ and\ \citenamefont
  {Lambiotte}}]{Schaub2017}%
  \BibitemOpen
  \bibfield  {author} {\bibinfo {author} {\bibfnamefont {M.~T.}\ \bibnamefont
  {Schaub}}, \bibinfo {author} {\bibfnamefont {J.-C.}\ \bibnamefont
  {Delvenne}}, \bibinfo {author} {\bibfnamefont {M.}~\bibnamefont {Rosvall}}, \
  and\ \bibinfo {author} {\bibfnamefont {R.}~\bibnamefont {Lambiotte}},\ }\href
  {\doibase 10.1007/s41109-017-0023-6} {\bibfield  {journal} {\bibinfo
  {journal} {Appl. Netw. Sci.}\ }\textbf {\bibinfo {volume} {2}},\ \bibinfo
  {pages} {4} (\bibinfo {year} {2017})}\BibitemShut {NoStop}%
\bibitem [{\citenamefont {Zhang}\ and\ \citenamefont
  {Moore}(2014)}]{Zhang2014b}%
  \BibitemOpen
  \bibfield  {author} {\bibinfo {author} {\bibfnamefont {P.}~\bibnamefont
  {Zhang}}\ and\ \bibinfo {author} {\bibfnamefont {C.}~\bibnamefont {Moore}},\
  }\href {\doibase 10.1073/pnas.1409770111} {\bibfield  {journal} {\bibinfo
  {journal} {Proc. Natl. Acad. Sci. U.S.A.}\ }\textbf {\bibinfo {volume}
  {111}},\ \bibinfo {pages} {18144} (\bibinfo {year} {2014})}\BibitemShut
  {NoStop}%
\bibitem [{\citenamefont {Newman}(2016)}]{newman2016equivalence}%
  \BibitemOpen
  \bibfield  {author} {\bibinfo {author} {\bibfnamefont {M.~E.~J.}\
  \bibnamefont {Newman}},\ }\href {\doibase 10.1103/PhysRevE.94.052315}
  {\bibfield  {journal} {\bibinfo  {journal} {Phys. Rev. E}\ }\textbf {\bibinfo
  {volume} {94}},\ \bibinfo {pages} {052315} (\bibinfo {year}
  {2016})}\BibitemShut {NoStop}%
\bibitem [{\citenamefont {Roxana~Pamfil}\ \emph {et~al.}(2018)\citenamefont
  {Roxana~Pamfil}, \citenamefont {Howison}, \citenamefont {Lambiotte},\ and\
  \citenamefont {Porter}}]{roxana2018relating}%
  \BibitemOpen
  \bibfield  {author} {\bibinfo {author} {\bibfnamefont {A.}~\bibnamefont
  {Roxana~Pamfil}}, \bibinfo {author} {\bibfnamefont {S.~D.}\ \bibnamefont
  {Howison}}, \bibinfo {author} {\bibfnamefont {R.}~\bibnamefont {Lambiotte}},
  \ and\ \bibinfo {author} {\bibfnamefont {M.~A.}\ \bibnamefont {Porter}},\
  }\href {https://arxiv.org/abs/1804.01964} {\bibfield  {journal} {\bibinfo
  {journal} {arXiv:1804.01964}\ } (\bibinfo {year} {2018})}\BibitemShut
  {NoStop}%
\bibitem [{\citenamefont {Veldt}\ \emph {et~al.}(2018)\citenamefont {Veldt},
  \citenamefont {Gleich},\ and\ \citenamefont {Wirth}}]{veldt2017unifying}%
  \BibitemOpen
  \bibfield  {author} {\bibinfo {author} {\bibfnamefont {N.}~\bibnamefont
  {Veldt}}, \bibinfo {author} {\bibfnamefont {D.}~\bibnamefont {Gleich}}, \
  and\ \bibinfo {author} {\bibfnamefont {A.}~\bibnamefont {Wirth}},\ }in\ \href
  {\doibase 10.1145/3178876.3186110} {\emph {\bibinfo {booktitle} {Proceedings
  of the 2018 World Wide Web Conference (WWW)}}}\ (\bibinfo {year} {2018})\
  pp.\ \bibinfo {pages} {439--448}\BibitemShut {NoStop}%
\bibitem [{\citenamefont {Newman}(2013{\natexlab{a}})}]{Newman2013}%
  \BibitemOpen
  \bibfield  {author} {\bibinfo {author} {\bibfnamefont {M.~E.~J.}\
  \bibnamefont {Newman}},\ }\href {\doibase 10.1103/PhysRevE.88.04282}
  {\bibfield  {journal} {\bibinfo  {journal} {Phys. Rev. E}\ }\textbf {\bibinfo
  {volume} {88}},\ \bibinfo {pages} {042822} (\bibinfo {year}
  {2013}{\natexlab{a}})}\BibitemShut {NoStop}%
\bibitem [{\citenamefont {Kawamoto}\ and\ \citenamefont
  {Kabashima}(2015)}]{kawamoto2015detectability}%
  \BibitemOpen
  \bibfield  {author} {\bibinfo {author} {\bibfnamefont {T.}~\bibnamefont
  {Kawamoto}}\ and\ \bibinfo {author} {\bibfnamefont {Y.}~\bibnamefont
  {Kabashima}},\ }\href {\doibase 10.1209/0295-5075/112/40007} {\bibfield
  {journal} {\bibinfo  {journal} {Europhys. Lett.}\ }\textbf {\bibinfo {volume}
  {112}},\ \bibinfo {pages} {40007} (\bibinfo {year} {2015})}\BibitemShut
  {NoStop}%
\bibitem [{\citenamefont {Masuda}\ \emph {et~al.}(2017)\citenamefont {Masuda},
  \citenamefont {Porter},\ and\ \citenamefont {Lambiotte}}]{masuda2017random}%
  \BibitemOpen
  \bibfield  {author} {\bibinfo {author} {\bibfnamefont {N.}~\bibnamefont
  {Masuda}}, \bibinfo {author} {\bibfnamefont {M.~A.}\ \bibnamefont {Porter}},
  \ and\ \bibinfo {author} {\bibfnamefont {R.}~\bibnamefont {Lambiotte}},\
  }\href {\doibase 10.1016/j.physrep.2017.07.007} {\bibfield  {journal}
  {\bibinfo  {journal} {Phys. Rep.}\ }\textbf {\bibinfo {volume} {716}}
  (\bibinfo {year} {2017}),\ 10.1016/j.physrep.2017.07.007}\BibitemShut
  {NoStop}%
\bibitem [{\citenamefont {Chang}\ \emph {et~al.}(2018)\citenamefont {Chang},
  \citenamefont {Cheng}, \citenamefont {Yan}, \citenamefont {Yin},\ and\
  \citenamefont {Zhang}}]{chang2018approximate}%
  \BibitemOpen
  \bibfield  {author} {\bibinfo {author} {\bibfnamefont {Z.}~\bibnamefont
  {Chang}}, \bibinfo {author} {\bibfnamefont {H.-M.}\ \bibnamefont {Cheng}},
  \bibinfo {author} {\bibfnamefont {C.}~\bibnamefont {Yan}}, \bibinfo {author}
  {\bibfnamefont {X.}~\bibnamefont {Yin}}, \ and\ \bibinfo {author}
  {\bibfnamefont {Z.-Y.}\ \bibnamefont {Zhang}},\ }\href
  {https://arxiv.org/abs/1801.03618} {\bibfield  {journal} {\bibinfo  {journal}
  {arXiv:1801.03618}\ } (\bibinfo {year} {2018})}\BibitemShut {NoStop}%
\bibitem [{\citenamefont {Karrer}\ and\ \citenamefont
  {Newman}(2011)}]{Karrer2011}%
  \BibitemOpen
  \bibfield  {author} {\bibinfo {author} {\bibfnamefont {B.}~\bibnamefont
  {Karrer}}\ and\ \bibinfo {author} {\bibfnamefont {M.~E.~J.}\ \bibnamefont
  {Newman}},\ }\href {\doibase 10.1103/PhysRevE.83.016107} {\bibfield
  {journal} {\bibinfo  {journal} {Phys. Rev. E}\ }\textbf {\bibinfo {volume}
  {83}},\ \bibinfo {pages} {016107} (\bibinfo {year} {2011})}\BibitemShut
  {NoStop}%
\bibitem [{\citenamefont {Ball}\ \emph {et~al.}(2011)\citenamefont {Ball},
  \citenamefont {Karrer},\ and\ \citenamefont {Newman}}]{Ball2011}%
  \BibitemOpen
  \bibfield  {author} {\bibinfo {author} {\bibfnamefont {B.}~\bibnamefont
  {Ball}}, \bibinfo {author} {\bibfnamefont {B.}~\bibnamefont {Karrer}}, \ and\
  \bibinfo {author} {\bibfnamefont {M.~E.~J.}\ \bibnamefont {Newman}},\ }\href
  {\doibase 10.1103/PhysRevE.84.036103} {\bibfield  {journal} {\bibinfo
  {journal} {Phys. Rev. E}\ }\textbf {\bibinfo {volume} {84}},\ \bibinfo
  {pages} {036103} (\bibinfo {year} {2011})}\BibitemShut {NoStop}%
\bibitem [{\citenamefont {Larremore}\ \emph {et~al.}(2014)\citenamefont
  {Larremore}, \citenamefont {Clauset},\ and\ \citenamefont
  {Jacobs}}]{larremore2014efficiently}%
  \BibitemOpen
  \bibfield  {author} {\bibinfo {author} {\bibfnamefont {D.~B.}\ \bibnamefont
  {Larremore}}, \bibinfo {author} {\bibfnamefont {A.}~\bibnamefont {Clauset}},
  \ and\ \bibinfo {author} {\bibfnamefont {A.~Z.}\ \bibnamefont {Jacobs}},\
  }\href {\doibase 10.1103/PhysRevE.90.012805} {\bibfield  {journal} {\bibinfo
  {journal} {Phys. Rev. E}\ }\textbf {\bibinfo {volume} {90}},\ \bibinfo
  {pages} {012805} (\bibinfo {year} {2014})}\BibitemShut {NoStop}%
\bibitem [{\citenamefont {Borgatti}\ and\ \citenamefont
  {Everett}(2000)}]{Borgatti2000}%
  \BibitemOpen
  \bibfield  {author} {\bibinfo {author} {\bibfnamefont {S.~P.}\ \bibnamefont
  {Borgatti}}\ and\ \bibinfo {author} {\bibfnamefont {M.~G.}\ \bibnamefont
  {Everett}},\ }\href {\doibase 10.1016/S0378-8733(99)00019-2} {\bibfield
  {journal} {\bibinfo  {journal} {Soc. Networks}\ }\textbf {\bibinfo {volume}
  {21}},\ \bibinfo {pages} {375} (\bibinfo {year} {2000})}\BibitemShut
  {NoStop}%
\bibitem [{\citenamefont {{\v{S}}ulc}\ and\ \citenamefont
  {Zdeborov{\'a}}(2010)}]{sulc2010belief}%
  \BibitemOpen
  \bibfield  {author} {\bibinfo {author} {\bibfnamefont {P.}~\bibnamefont
  {{\v{S}}ulc}}\ and\ \bibinfo {author} {\bibfnamefont {L.}~\bibnamefont
  {Zdeborov{\'a}}},\ }\href {\doibase 10.1088/1751-8113/43/28/285003}
  {\bibfield  {journal} {\bibinfo  {journal} {J. Phys. A}\ }\textbf {\bibinfo
  {volume} {43}},\ \bibinfo {pages} {285003} (\bibinfo {year}
  {2010})}\BibitemShut {NoStop}%
\bibitem [{\citenamefont {Hric}\ \emph {et~al.}(2016)\citenamefont {Hric},
  \citenamefont {Peixoto},\ and\ \citenamefont {Fortunato}}]{hric2016network}%
  \BibitemOpen
  \bibfield  {author} {\bibinfo {author} {\bibfnamefont {D.}~\bibnamefont
  {Hric}}, \bibinfo {author} {\bibfnamefont {T.~P.}\ \bibnamefont {Peixoto}}, \
  and\ \bibinfo {author} {\bibfnamefont {S.}~\bibnamefont {Fortunato}},\ }\href
  {\doibase 10.1103/PhysRevX.6.031038} {\bibfield  {journal} {\bibinfo
  {journal} {Phys. Rev. X}\ }\textbf {\bibinfo {volume} {6}},\ \bibinfo {pages}
  {031038} (\bibinfo {year} {2016})}\BibitemShut {NoStop}%
\bibitem [{\citenamefont {Newman}\ and\ \citenamefont
  {Clauset}(2016)}]{newman2016structure}%
  \BibitemOpen
  \bibfield  {author} {\bibinfo {author} {\bibfnamefont {M.~E.~J.}\
  \bibnamefont {Newman}}\ and\ \bibinfo {author} {\bibfnamefont
  {A.}~\bibnamefont {Clauset}},\ }\href {\doibase 10.1038/ncomms11863}
  {\bibfield  {journal} {\bibinfo  {journal} {Nat. Comm.}\ }\textbf {\bibinfo
  {volume} {7}},\ \bibinfo {pages} {11863} (\bibinfo {year}
  {2016})}\BibitemShut {NoStop}%
\bibitem [{\citenamefont {Arenas}\ \emph {et~al.}(2007)\citenamefont {Arenas},
  \citenamefont {Duch}, \citenamefont {Fern{\'a}ndez},\ and\ \citenamefont
  {G{\'o}mez}}]{arenas2007size}%
  \BibitemOpen
  \bibfield  {author} {\bibinfo {author} {\bibfnamefont {A.}~\bibnamefont
  {Arenas}}, \bibinfo {author} {\bibfnamefont {J.}~\bibnamefont {Duch}},
  \bibinfo {author} {\bibfnamefont {A.}~\bibnamefont {Fern{\'a}ndez}}, \ and\
  \bibinfo {author} {\bibfnamefont {S.}~\bibnamefont {G{\'o}mez}},\ }\href
  {\doibase 10.1088/1367-2630/9/6/176} {\bibfield  {journal} {\bibinfo
  {journal} {New J. Phys.}\ }\textbf {\bibinfo {volume} {9}},\ \bibinfo {pages}
  {176} (\bibinfo {year} {2007})}\BibitemShut {NoStop}%
\bibitem [{\citenamefont {Clauset}\ \emph {et~al.}(2008)\citenamefont
  {Clauset}, \citenamefont {Moore},\ and\ \citenamefont
  {Newman}}]{clauset2008hierarchical}%
  \BibitemOpen
  \bibfield  {author} {\bibinfo {author} {\bibfnamefont {A.}~\bibnamefont
  {Clauset}}, \bibinfo {author} {\bibfnamefont {C.}~\bibnamefont {Moore}}, \
  and\ \bibinfo {author} {\bibfnamefont {M.~E.~J.}\ \bibnamefont {Newman}},\
  }\href {\doibase 10.1038/nature06830} {\bibfield  {journal} {\bibinfo
  {journal} {Nature}\ }\textbf {\bibinfo {volume} {453}},\ \bibinfo {pages}
  {98} (\bibinfo {year} {2008})}\BibitemShut {NoStop}%
\bibitem [{\citenamefont {Blondel}\ \emph {et~al.}(2008)\citenamefont
  {Blondel}, \citenamefont {Guillaume}, \citenamefont {Lambiotte},\ and\
  \citenamefont {Lefebvre}}]{blondel2008fast}%
  \BibitemOpen
  \bibfield  {author} {\bibinfo {author} {\bibfnamefont {V.~D.}\ \bibnamefont
  {Blondel}}, \bibinfo {author} {\bibfnamefont {J.-L.}\ \bibnamefont
  {Guillaume}}, \bibinfo {author} {\bibfnamefont {R.}~\bibnamefont
  {Lambiotte}}, \ and\ \bibinfo {author} {\bibfnamefont {E.}~\bibnamefont
  {Lefebvre}},\ }\href {\doibase 10.1088/1742-5468/2008/10/P10008} {\bibfield
  {journal} {\bibinfo  {journal} {J. Stat. Mech. Theor. Exp.}\ }\textbf
  {\bibinfo {volume} {2008}},\ \bibinfo {pages} {P10008} (\bibinfo {year}
  {2008})}\BibitemShut {NoStop}%
\bibitem [{\citenamefont {Fortunato}(2010)}]{Fortunato2010}%
  \BibitemOpen
  \bibfield  {author} {\bibinfo {author} {\bibfnamefont {S.}~\bibnamefont
  {Fortunato}},\ }\href {\doibase 10.1016/j.physrep.2009.11.002} {\bibfield
  {journal} {\bibinfo  {journal} {Phys. Rep.}\ }\textbf {\bibinfo {volume}
  {486}},\ \bibinfo {pages} {75} (\bibinfo {year} {2010})}\BibitemShut
  {NoStop}%
\bibitem [{\citenamefont {Porter}\ \emph {et~al.}(2009)\citenamefont {Porter},
  \citenamefont {Onnela},\ and\ \citenamefont {Mucha}}]{Porter2009}%
  \BibitemOpen
  \bibfield  {author} {\bibinfo {author} {\bibfnamefont {M.~A.}\ \bibnamefont
  {Porter}}, \bibinfo {author} {\bibfnamefont {J.-P.}\ \bibnamefont {Onnela}},
  \ and\ \bibinfo {author} {\bibfnamefont {P.~J.}\ \bibnamefont {Mucha}},\
  }\href {http://www.unc.edu/~mucha/Reprints/NoticesCommunities.pdf} {\bibfield
   {journal} {\bibinfo  {journal} {Notices of the AMS}\ }\textbf {\bibinfo
  {volume} {56}},\ \bibinfo {pages} {1082} (\bibinfo {year}
  {2009})}\BibitemShut {NoStop}%
\bibitem [{\citenamefont {Shai}\ \emph {et~al.}(2017)\citenamefont {Shai},
  \citenamefont {Stanley}, \citenamefont {Granell}, \citenamefont {Taylor},\
  and\ \citenamefont {Mucha}}]{shai2017case}%
  \BibitemOpen
  \bibfield  {author} {\bibinfo {author} {\bibfnamefont {S.}~\bibnamefont
  {Shai}}, \bibinfo {author} {\bibfnamefont {N.}~\bibnamefont {Stanley}},
  \bibinfo {author} {\bibfnamefont {C.}~\bibnamefont {Granell}}, \bibinfo
  {author} {\bibfnamefont {D.}~\bibnamefont {Taylor}}, \ and\ \bibinfo {author}
  {\bibfnamefont {P.~J.}\ \bibnamefont {Mucha}},\ }\href
  {https://arxiv.org/abs/1705.02305} {\bibfield  {journal} {\bibinfo  {journal}
  {arXiv:1705.02305}\ } (\bibinfo {year} {2017})}\BibitemShut {NoStop}%
\bibitem [{\citenamefont {Newman}\ and\ \citenamefont
  {Girvan}(2004)}]{Newman2004}%
  \BibitemOpen
  \bibfield  {author} {\bibinfo {author} {\bibfnamefont {M.~E.~J.}\
  \bibnamefont {Newman}}\ and\ \bibinfo {author} {\bibfnamefont
  {M.}~\bibnamefont {Girvan}},\ }\href {\doibase 10.1103/PhysRevE.69.026113}
  {\bibfield  {journal} {\bibinfo  {journal} {Phys. Rev. E}\ }\textbf {\bibinfo
  {volume} {69}},\ \bibinfo {pages} {026113} (\bibinfo {year}
  {2004})}\BibitemShut {NoStop}%
\bibitem [{\citenamefont {Rombach}\ \emph {et~al.}(2014)\citenamefont
  {Rombach}, \citenamefont {Porter}, \citenamefont {Fowler},\ and\
  \citenamefont {Mucha}}]{Rombach2014}%
  \BibitemOpen
  \bibfield  {author} {\bibinfo {author} {\bibfnamefont {M.~P.}\ \bibnamefont
  {Rombach}}, \bibinfo {author} {\bibfnamefont {M.~A.}\ \bibnamefont {Porter}},
  \bibinfo {author} {\bibfnamefont {J.~H.}\ \bibnamefont {Fowler}}, \ and\
  \bibinfo {author} {\bibfnamefont {P.~J.}\ \bibnamefont {Mucha}},\ }\href
  {\doibase 10.1137/120881683} {\bibfield  {journal} {\bibinfo  {journal} {SIAM
  J. Appl. Math.}\ }\textbf {\bibinfo {volume} {74}},\ \bibinfo {pages} {167}
  (\bibinfo {year} {2014})}\BibitemShut {NoStop}%
\bibitem [{\citenamefont {Brandes}\ \emph {et~al.}(2006)\citenamefont
  {Brandes}, \citenamefont {Delling}, \citenamefont {Gaertler}, \citenamefont
  {G{\"o}rke}, \citenamefont {Hoefer}, \citenamefont {Nikoloski},\ and\
  \citenamefont {Wagner}}]{Brandes2006}%
  \BibitemOpen
  \bibfield  {author} {\bibinfo {author} {\bibfnamefont {U.}~\bibnamefont
  {Brandes}}, \bibinfo {author} {\bibfnamefont {D.}~\bibnamefont {Delling}},
  \bibinfo {author} {\bibfnamefont {M.}~\bibnamefont {Gaertler}}, \bibinfo
  {author} {\bibfnamefont {R.}~\bibnamefont {G{\"o}rke}}, \bibinfo {author}
  {\bibfnamefont {M.}~\bibnamefont {Hoefer}}, \bibinfo {author} {\bibfnamefont
  {Z.}~\bibnamefont {Nikoloski}}, \ and\ \bibinfo {author} {\bibfnamefont
  {D.}~\bibnamefont {Wagner}},\ }\href {https://arxiv.org/abs/physics/0608255}
  {\bibfield  {journal} {\bibinfo  {journal} {arXiv:physics/0608255}\ }
  (\bibinfo {year} {2006})}\BibitemShut {NoStop}%
\bibitem [{\citenamefont {Crescenzi}\ and\ \citenamefont
  {Kann}(1997)}]{Crescenzi1995}%
  \BibitemOpen
  \bibfield  {author} {\bibinfo {author} {\bibfnamefont {P.}~\bibnamefont
  {Crescenzi}}\ and\ \bibinfo {author} {\bibfnamefont {V.}~\bibnamefont
  {Kann}},\ }\href {\doibase 10.1007/3-540-63248-4_10} {\emph {\bibinfo {title}
  {International Workshop on Randomization and Approximation Techniques in
  Computer Science}}}\ (\bibinfo {year} {1997})\ pp.\ \bibinfo {pages}
  {111--118}\BibitemShut {NoStop}%
\bibitem [{\citenamefont {Stein}\ and\ \citenamefont
  {Newman}(2013)}]{Stein2013}%
  \BibitemOpen
  \bibfield  {author} {\bibinfo {author} {\bibfnamefont {D.~L.}\ \bibnamefont
  {Stein}}\ and\ \bibinfo {author} {\bibfnamefont {C.~M.}\ \bibnamefont
  {Newman}},\ }\href@noop {} {\emph {\bibinfo {title} {{Spin Glasses and
  Complexity}}}}\ (\bibinfo  {publisher} {Princeton University Press,
  Princeton, NJ},\ \bibinfo {year} {2013})\BibitemShut {NoStop}%
\bibitem [{\citenamefont {Peel}\ \emph {et~al.}(2017)\citenamefont {Peel},
  \citenamefont {Larremore},\ and\ \citenamefont {Clauset}}]{Peel2017}%
  \BibitemOpen
  \bibfield  {author} {\bibinfo {author} {\bibfnamefont {L.}~\bibnamefont
  {Peel}}, \bibinfo {author} {\bibfnamefont {D.~B.}\ \bibnamefont {Larremore}},
  \ and\ \bibinfo {author} {\bibfnamefont {A.}~\bibnamefont {Clauset}},\ }\href
  {\doibase 10.1126/sciadv.1602548} {\bibfield  {journal} {\bibinfo  {journal}
  {Sci. Adv.}\ }\textbf {\bibinfo {volume} {3}},\ \bibinfo {pages} {e1602548}
  (\bibinfo {year} {2017})}\BibitemShut {NoStop}%
\bibitem [{\citenamefont {Kernighan}\ and\ \citenamefont
  {Lin}(1970)}]{kernighan1970efficient}%
  \BibitemOpen
  \bibfield  {author} {\bibinfo {author} {\bibfnamefont {B.~W.}\ \bibnamefont
  {Kernighan}}\ and\ \bibinfo {author} {\bibfnamefont {S.}~\bibnamefont
  {Lin}},\ }\href {\doibase 10.1002/j.1538-7305.1970.tb01770.x} {\bibfield
  {journal} {\bibinfo  {journal} {Bell Syst. Tech. J}\ }\textbf {\bibinfo
  {volume} {49}},\ \bibinfo {pages} {291} (\bibinfo {year} {1970})}\BibitemShut
  {NoStop}%
\bibitem [{\citenamefont {Newman}(2006)}]{Newman2006}%
  \BibitemOpen
  \bibfield  {author} {\bibinfo {author} {\bibfnamefont {M.~E.~J.}\
  \bibnamefont {Newman}},\ }\href {\doibase 10.1073/pnas.0601602103} {\bibfield
   {journal} {\bibinfo  {journal} {Proc. Natl. Acad. Sci. U.S.A.}\ }\textbf
  {\bibinfo {volume} {103}},\ \bibinfo {pages} {8577} (\bibinfo {year}
  {2006})}\BibitemShut {NoStop}%
\bibitem [{\citenamefont {White}\ and\ \citenamefont
  {Smyth}(2005)}]{white2005spectral}%
  \BibitemOpen
  \bibfield  {author} {\bibinfo {author} {\bibfnamefont {S.}~\bibnamefont
  {White}}\ and\ \bibinfo {author} {\bibfnamefont {P.}~\bibnamefont {Smyth}},\
  }in\ \href {http://www.datalab.uci.edu/papers/siam_graph_clustering.pdf}
  {\emph {\bibinfo {booktitle} {Proceedings of the 2005 {SIAM} {International}
  Conference on Data Mining}}}\ (\bibinfo {organization} {SIAM, Philadelphia,
  PA},\ \bibinfo {year} {2005})\ pp.\ \bibinfo {pages} {274--285}\BibitemShut
  {NoStop}%
\bibitem [{\citenamefont {Raghavan}\ \emph {et~al.}(2007)\citenamefont
  {Raghavan}, \citenamefont {Albert},\ and\ \citenamefont
  {Kumara}}]{raghavan2007near}%
  \BibitemOpen
  \bibfield  {author} {\bibinfo {author} {\bibfnamefont {U.~N.}\ \bibnamefont
  {Raghavan}}, \bibinfo {author} {\bibfnamefont {R.}~\bibnamefont {Albert}}, \
  and\ \bibinfo {author} {\bibfnamefont {S.}~\bibnamefont {Kumara}},\ }\href
  {\doibase 10.1103/PhysRevE.76.036106} {\bibfield  {journal} {\bibinfo
  {journal} {Phys. Rev. E}\ }\textbf {\bibinfo {volume} {76}},\ \bibinfo
  {pages} {036106} (\bibinfo {year} {2007})}\BibitemShut {NoStop}%
\bibitem [{\citenamefont {Tib{\'e}ly}\ and\ \citenamefont
  {Kert{\'e}sz}(2008)}]{tibely2008equivalence}%
  \BibitemOpen
  \bibfield  {author} {\bibinfo {author} {\bibfnamefont {G.}~\bibnamefont
  {Tib{\'e}ly}}\ and\ \bibinfo {author} {\bibfnamefont {J.}~\bibnamefont
  {Kert{\'e}sz}},\ }\href {\doibase 10.1016/j.physa.2008.04.024} {\bibfield
  {journal} {\bibinfo  {journal} {Physica A}\ }\textbf {\bibinfo {volume}
  {387}},\ \bibinfo {pages} {4982} (\bibinfo {year} {2008})}\BibitemShut
  {NoStop}%
\bibitem [{\citenamefont {Barber}\ and\ \citenamefont
  {Clark}(2009)}]{barber2009detecting}%
  \BibitemOpen
  \bibfield  {author} {\bibinfo {author} {\bibfnamefont {M.~J.}\ \bibnamefont
  {Barber}}\ and\ \bibinfo {author} {\bibfnamefont {J.~W.}\ \bibnamefont
  {Clark}},\ }\href {\doibase 10.1103/PhysRevE.80.026129} {\bibfield  {journal}
  {\bibinfo  {journal} {Phys. Rev. E}\ }\textbf {\bibinfo {volume} {80}},\
  \bibinfo {pages} {026129} (\bibinfo {year} {2009})}\BibitemShut {NoStop}%
\bibitem [{\citenamefont {Ver~Steeg}\ \emph {et~al.}(2014)\citenamefont
  {Ver~Steeg}, \citenamefont {Moore}, \citenamefont {Galstyan},\ and\
  \citenamefont {Allahverdyan}}]{ver2014phase}%
  \BibitemOpen
  \bibfield  {author} {\bibinfo {author} {\bibfnamefont {G.}~\bibnamefont
  {Ver~Steeg}}, \bibinfo {author} {\bibfnamefont {C.}~\bibnamefont {Moore}},
  \bibinfo {author} {\bibfnamefont {A.}~\bibnamefont {Galstyan}}, \ and\
  \bibinfo {author} {\bibfnamefont {A.}~\bibnamefont {Allahverdyan}},\ }\href
  {\doibase 10.1209/0295-5075/106/48004} {\bibfield  {journal} {\bibinfo
  {journal} {Europhys. Lett.}\ }\textbf {\bibinfo {volume} {106}},\ \bibinfo
  {pages} {48004} (\bibinfo {year} {2014})}\BibitemShut {NoStop}%
\bibitem [{\citenamefont {Reichardt}\ and\ \citenamefont
  {Leone}(2008)}]{Reichardt2008}%
  \BibitemOpen
  \bibfield  {author} {\bibinfo {author} {\bibfnamefont {J.}~\bibnamefont
  {Reichardt}}\ and\ \bibinfo {author} {\bibfnamefont {M.}~\bibnamefont
  {Leone}},\ }\href {\doibase 10.1103/PhysRevLett.101.078701} {\bibfield
  {journal} {\bibinfo  {journal} {Phys. Rev. Lett.}\ }\textbf {\bibinfo
  {volume} {101}},\ \bibinfo {pages} {078701} (\bibinfo {year}
  {2008})}\BibitemShut {NoStop}%
\bibitem [{\citenamefont {Reichardt}\ and\ \citenamefont
  {Bornholdt}(2006)}]{Reichardt2006}%
  \BibitemOpen
  \bibfield  {author} {\bibinfo {author} {\bibfnamefont {J.}~\bibnamefont
  {Reichardt}}\ and\ \bibinfo {author} {\bibfnamefont {S.}~\bibnamefont
  {Bornholdt}},\ }\href {\doibase 10.1103/PhysRevE.74.016110} {\bibfield
  {journal} {\bibinfo  {journal} {Phys. Rev. E}\ }\textbf {\bibinfo {volume}
  {74}},\ \bibinfo {pages} {016110} (\bibinfo {year} {2006})}\BibitemShut
  {NoStop}%
\bibitem [{\citenamefont {Decelle}\ \emph {et~al.}(2011)\citenamefont
  {Decelle}, \citenamefont {Krzakala}, \citenamefont {Moore},\ and\
  \citenamefont {Zdeborov{\'a}}}]{Decelle2011a}%
  \BibitemOpen
  \bibfield  {author} {\bibinfo {author} {\bibfnamefont {A.}~\bibnamefont
  {Decelle}}, \bibinfo {author} {\bibfnamefont {F.}~\bibnamefont {Krzakala}},
  \bibinfo {author} {\bibfnamefont {C.}~\bibnamefont {Moore}}, \ and\ \bibinfo
  {author} {\bibfnamefont {L.}~\bibnamefont {Zdeborov{\'a}}},\ }\href {\doibase
  10.1103/PhysRevLett.107.065701} {\bibfield  {journal} {\bibinfo  {journal}
  {Phys. Rev. Lett.}\ }\textbf {\bibinfo {volume} {107}},\ \bibinfo {pages}
  {065701} (\bibinfo {year} {2011})}\BibitemShut {NoStop}%
\bibitem [{\citenamefont {Ronhovde}\ and\ \citenamefont
  {Nussinov}(2010)}]{ronhovde2010local}%
  \BibitemOpen
  \bibfield  {author} {\bibinfo {author} {\bibfnamefont {P.}~\bibnamefont
  {Ronhovde}}\ and\ \bibinfo {author} {\bibfnamefont {Z.}~\bibnamefont
  {Nussinov}},\ }\href {\doibase 10.1103/PhysRevE.81.046114} {\bibfield
  {journal} {\bibinfo  {journal} {Phys. Rev. E}\ }\textbf {\bibinfo {volume}
  {81}},\ \bibinfo {pages} {046114} (\bibinfo {year} {2010})}\BibitemShut
  {NoStop}%
\bibitem [{\citenamefont {Kirkpatrick}\ \emph {et~al.}(1983)\citenamefont
  {Kirkpatrick}, \citenamefont {Gelatt},\ and\ \citenamefont
  {Vecchi}}]{kirkpatrick1983optimization}%
  \BibitemOpen
  \bibfield  {author} {\bibinfo {author} {\bibfnamefont {S.}~\bibnamefont
  {Kirkpatrick}}, \bibinfo {author} {\bibfnamefont {C.~D.}\ \bibnamefont
  {Gelatt}}, \ and\ \bibinfo {author} {\bibfnamefont {M.~P.}\ \bibnamefont
  {Vecchi}},\ }\href {\doibase 10.1126/science.220.4598.671} {\bibfield
  {journal} {\bibinfo  {journal} {Science}\ }\textbf {\bibinfo {volume}
  {220}},\ \bibinfo {pages} {671} (\bibinfo {year} {1983})}\BibitemShut
  {NoStop}%
\bibitem [{\citenamefont {Holland}\ \emph {et~al.}(1983)\citenamefont
  {Holland}, \citenamefont {Laskey},\ and\ \citenamefont
  {Leinhardt}}]{Holland1983}%
  \BibitemOpen
  \bibfield  {author} {\bibinfo {author} {\bibfnamefont {P.~W.}\ \bibnamefont
  {Holland}}, \bibinfo {author} {\bibfnamefont {K.~B.}\ \bibnamefont {Laskey}},
  \ and\ \bibinfo {author} {\bibfnamefont {S.}~\bibnamefont {Leinhardt}},\
  }\href {\doibase 10.1016/0378-8733(83)90021-7} {\bibfield  {journal}
  {\bibinfo  {journal} {Soc. Networks}\ }\textbf {\bibinfo {volume} {5}},\
  \bibinfo {pages} {109} (\bibinfo {year} {1983})}\BibitemShut {NoStop}%
\bibitem [{\citenamefont {Kawamoto}\ and\ \citenamefont
  {Kabashima}(2017)}]{Kawamoto2017}%
  \BibitemOpen
  \bibfield  {author} {\bibinfo {author} {\bibfnamefont {T.}~\bibnamefont
  {Kawamoto}}\ and\ \bibinfo {author} {\bibfnamefont {Y.}~\bibnamefont
  {Kabashima}},\ }\href {\doibase 10.1103/PhysRevE.95.012304} {\bibfield
  {journal} {\bibinfo  {journal} {Phys. Rev. E}\ }\textbf {\bibinfo {volume}
  {95}},\ \bibinfo {pages} {012304} (\bibinfo {year} {2017})}\BibitemShut
  {NoStop}%
\bibitem [{\citenamefont {Young}\ \emph {et~al.}(2017)\citenamefont {Young},
  \citenamefont {Desrosiers}, \citenamefont {H{\'e}bert-Dufresne},
  \citenamefont {Laurence},\ and\ \citenamefont {Dub{\'e}}}]{Young2017}%
  \BibitemOpen
  \bibfield  {author} {\bibinfo {author} {\bibfnamefont {J.-G.}\ \bibnamefont
  {Young}}, \bibinfo {author} {\bibfnamefont {P.}~\bibnamefont {Desrosiers}},
  \bibinfo {author} {\bibfnamefont {L.}~\bibnamefont {H{\'e}bert-Dufresne}},
  \bibinfo {author} {\bibfnamefont {E.}~\bibnamefont {Laurence}}, \ and\
  \bibinfo {author} {\bibfnamefont {L.~J.}\ \bibnamefont {Dub{\'e}}},\ }\href
  {\doibase 10.1103/PhysRevE.95.062304} {\bibfield  {journal} {\bibinfo
  {journal} {Phys. Rev. E}\ }\textbf {\bibinfo {volume} {95}},\ \bibinfo
  {pages} {062304} (\bibinfo {year} {2017})}\BibitemShut {NoStop}%
\bibitem [{\citenamefont {Newman}(2013{\natexlab{b}})}]{newman2013community}%
  \BibitemOpen
  \bibfield  {author} {\bibinfo {author} {\bibfnamefont {M.~E.~J.}\
  \bibnamefont {Newman}},\ }\href {\doibase 10.1209/0295-5075/103/28003}
  {\bibfield  {journal} {\bibinfo  {journal} {Europhys. Lett.}\ }\textbf
  {\bibinfo {volume} {103}},\ \bibinfo {pages} {28003} (\bibinfo {year}
  {2013}{\natexlab{b}})}\BibitemShut {NoStop}%
\bibitem [{\citenamefont {Krzakala}\ and\ \citenamefont
  {Zdeborov{\'a}}(2009)}]{krzakala2009hiding}%
  \BibitemOpen
  \bibfield  {author} {\bibinfo {author} {\bibfnamefont {F.}~\bibnamefont
  {Krzakala}}\ and\ \bibinfo {author} {\bibfnamefont {L.}~\bibnamefont
  {Zdeborov{\'a}}},\ }\href {\doibase 10.1103/PhysRevLett.102.238701}
  {\bibfield  {journal} {\bibinfo  {journal} {Phys. Rev. Lett.}\ }\textbf
  {\bibinfo {volume} {102}},\ \bibinfo {pages} {238701} (\bibinfo {year}
  {2009})}\BibitemShut {NoStop}%
\bibitem [{\citenamefont {Kojaku}\ and\ \citenamefont
  {Masuda}(2017)}]{kojaku2017finding}%
  \BibitemOpen
  \bibfield  {author} {\bibinfo {author} {\bibfnamefont {S.}~\bibnamefont
  {Kojaku}}\ and\ \bibinfo {author} {\bibfnamefont {N.}~\bibnamefont
  {Masuda}},\ }\href {\doibase PhysRevE.96.052313} {\bibfield  {journal}
  {\bibinfo  {journal} {Phys. Rev. E}\ }\textbf {\bibinfo {volume} {96}},\
  \bibinfo {pages} {052313} (\bibinfo {year} {2017})}\BibitemShut {NoStop}%
\bibitem [{\citenamefont {Traag}\ \emph {et~al.}(2011)\citenamefont {Traag},
  \citenamefont {Van~Dooren},\ and\ \citenamefont {Nesterov}}]{Traag2011}%
  \BibitemOpen
  \bibfield  {author} {\bibinfo {author} {\bibfnamefont {V.~A.}\ \bibnamefont
  {Traag}}, \bibinfo {author} {\bibfnamefont {P.}~\bibnamefont {Van~Dooren}}, \
  and\ \bibinfo {author} {\bibfnamefont {Y.}~\bibnamefont {Nesterov}},\ }\href
  {\doibase 10.1103/PhysRevE.84.016114} {\bibfield  {journal} {\bibinfo
  {journal} {Phys. Rev. E}\ }\textbf {\bibinfo {volume} {84}},\ \bibinfo
  {pages} {016114} (\bibinfo {year} {2011})}\BibitemShut {NoStop}%
\bibitem [{\citenamefont {Expert}\ \emph {et~al.}(2011)\citenamefont {Expert},
  \citenamefont {Evans}, \citenamefont {Blondel},\ and\ \citenamefont
  {Lambiotte}}]{expert2011uncovering}%
  \BibitemOpen
  \bibfield  {author} {\bibinfo {author} {\bibfnamefont {P.}~\bibnamefont
  {Expert}}, \bibinfo {author} {\bibfnamefont {T.~S.}\ \bibnamefont {Evans}},
  \bibinfo {author} {\bibfnamefont {V.~D.}\ \bibnamefont {Blondel}}, \ and\
  \bibinfo {author} {\bibfnamefont {R.}~\bibnamefont {Lambiotte}},\ }\href
  {\doibase 10.1073/pnas.1018962108} {\bibfield  {journal} {\bibinfo  {journal}
  {Proc. Natl Acad. of Sci.}\ }\textbf {\bibinfo {volume} {108}},\ \bibinfo
  {pages} {7663} (\bibinfo {year} {2011})}\BibitemShut {NoStop}%
\bibitem [{\citenamefont {Olhede}\ and\ \citenamefont
  {Wolfe}(2014)}]{olhede2014network}%
  \BibitemOpen
  \bibfield  {author} {\bibinfo {author} {\bibfnamefont {S.~C.}\ \bibnamefont
  {Olhede}}\ and\ \bibinfo {author} {\bibfnamefont {P.~J.}\ \bibnamefont
  {Wolfe}},\ }\href {\doibase 10.1073/pnas.1400374111} {\bibfield  {journal}
  {\bibinfo  {journal} {Proc. Natl Acad. of Sci.}\ }\textbf {\bibinfo {volume}
  {111}},\ \bibinfo {pages} {14722} (\bibinfo {year} {2014})}\BibitemShut
  {NoStop}%
\bibitem [{\citenamefont {Peixoto}(2014)}]{Peixoto2014}%
  \BibitemOpen
  \bibfield  {author} {\bibinfo {author} {\bibfnamefont {T.~P.}\ \bibnamefont
  {Peixoto}},\ }\href {\doibase 10.1103/PhysRevE.89.012804} {\bibfield
  {journal} {\bibinfo  {journal} {Phys. Rev. E}\ }\textbf {\bibinfo {volume}
  {89}},\ \bibinfo {pages} {012804} (\bibinfo {year} {2014})}\BibitemShut
  {NoStop}%
\bibitem [{\citenamefont {Kawamoto}(2017)}]{kawamoto2017algorithmic}%
  \BibitemOpen
  \bibfield  {author} {\bibinfo {author} {\bibfnamefont {T.}~\bibnamefont
  {Kawamoto}},\ }\href {https://arxiv.org/abs/1710.08816} {\bibfield  {journal}
  {\bibinfo  {journal} {arXiv:1710.08816}\ } (\bibinfo {year}
  {2017})}\BibitemShut {NoStop}%
\bibitem [{\citenamefont {Kawamoto}(2018)}]{kawamoto2018algorithmic}%
  \BibitemOpen
  \bibfield  {author} {\bibinfo {author} {\bibfnamefont {T.}~\bibnamefont
  {Kawamoto}},\ }\href {\doibase 10.1103/Phys-RevE.97.032301} {\bibfield
  {journal} {\bibinfo  {journal} {Phys. Rev. E}\ }\textbf {\bibinfo {volume}
  {97}},\ \bibinfo {pages} {032301} (\bibinfo {year} {2018})}\BibitemShut
  {NoStop}%
\bibitem [{\citenamefont {Ghasemian}\ \emph {et~al.}(2018)\citenamefont
  {Ghasemian}, \citenamefont {Hosseinmardi},\ and\ \citenamefont
  {Clauset}}]{ghasemian2018evaluating}%
  \BibitemOpen
  \bibfield  {author} {\bibinfo {author} {\bibfnamefont {A.}~\bibnamefont
  {Ghasemian}}, \bibinfo {author} {\bibfnamefont {H.}~\bibnamefont
  {Hosseinmardi}}, \ and\ \bibinfo {author} {\bibfnamefont {A.}~\bibnamefont
  {Clauset}},\ }\href {https://arxiv.org/abs/1802.10582} {\bibfield  {journal}
  {\bibinfo  {journal} {arXiv:1802.10582}\ } (\bibinfo {year}
  {2018})}\BibitemShut {NoStop}%
\bibitem [{\citenamefont {Kawamoto}\ and\ \citenamefont
  {Kabashima}(2018)}]{kawamoto2018comparative}%
  \BibitemOpen
  \bibfield  {author} {\bibinfo {author} {\bibfnamefont {T.}~\bibnamefont
  {Kawamoto}}\ and\ \bibinfo {author} {\bibfnamefont {Y.}~\bibnamefont
  {Kabashima}},\ }\href {\doibase 10.1103/PhysRevE.97.022315} {\bibfield
  {journal} {\bibinfo  {journal} {Phys. Rev. E}\ }\textbf {\bibinfo {volume}
  {97}},\ \bibinfo {pages} {022315} (\bibinfo {year} {2018})}\BibitemShut
  {NoStop}%
\bibitem [{\citenamefont {Abbe}(2018)}]{abbe2018community}%
  \BibitemOpen
  \bibfield  {author} {\bibinfo {author} {\bibfnamefont {E.}~\bibnamefont
  {Abbe}},\ }\href {http://jmlr.org/papers/v18/16-480.html} {\bibfield
  {journal} {\bibinfo  {journal} {Journal of Machine Learning Research}\
  }\textbf {\bibinfo {volume} {18}},\ \bibinfo {pages} {1} (\bibinfo {year}
  {2018})}\BibitemShut {NoStop}%
\end{thebibliography}
\bibliographystyle{apsrev4-1}

\end{document}